# Disorder Modifications of the Critical Temperature for Superconductivity – A Perspective from the Point of View of Nanoscience

Clifford M. Krowne

Electromagnetics Technology Branch, Electronics Science and Technology Division
Naval Research Laboratory, Washington, DC 20375

## Abstract

Superconductivity can be modified by various effects related to randomness, disorder, structural defects, and other similar physical effects. Their affects on superconductivity are important because such effects are intrinsic to certain material system's preparation, or may be intentionally produced. In this chapter, we show in the context of a Cooper instability relationship, that introduction of disorder through impurities could possibly lead to an increase in $T_c$. This is an old subject, having been addressed decades ago in the view of simpler substances, including metal alloy materials. Today, with the advances in material science, nanoscience, and atomic level preparation of materials and devices, this subject should be reexamined. That is our purpose here, especially in light of some recent discoveries made in the area of the metal-insulator transition, to be covered herein. One may obtain a formula for $T_c$ which is recognizable, and relatable to BCS theory, by using many-body quantum field theory for condensed matter expeditiously, using quantum Green's functions, polarization functions (correlation functions), and vertex functions, to extract out an analytical formula for $T_c$. The reader will find use made of perturbational Feynman diagrammatic techniques, finite temperature quantum Matsubara Green's functions, and quantum perturbational derivations without the diagrams.

# Disorder Modifications of the Critical Temperature for Superconductivity – A Perspective from the Point of View of Nanoscience

Clifford M. Krowne

Electromagnetics Technology Branch, Electronics Science and Technology Division
Naval Research Laboratory, Washington, DC 20375

## I. INTRODUCTION

Superconductivity can be modified by various effects related to randomness, disorder, structural defects, and other similar physical effects. Their affects on superconductivity are important because such effects are intrinsic to certain material system's preparation, or may be intentionally produced. In this chapter, we show in the context of a Cooper instability relationship, that introduction of disorder through impurities could possibly lead to an increase in $T_c$. This is an old subject, having been addressed decades ago in the view of simpler substances, including metal alloy materials. Today, with the advances in material science, nanoscience, and atomic level preparation of materials and devices, this subject should be reexamined. That is our purpose here, especially in light of some recent discoveries made in the area of the metal-insulator transition, to be covered below shortly. Although there is a complicated collection of quantum Green's functions, polarization functions (correlation functions), and vertex functions, used in the hierarchy of assumptions to extract out an analytical formula for $T_c$, one may obtain a formula for it which is recognizable, and relatable to BCS theory, by using many-body quantum field theory for condensed matter expeditiously. Handling modified BCS superconductors, medium temperature superconductors, and high $T_c$ superconductors may be possible by appropriate use of more accurate function representations and appropriate adaptions. All of this analysis is done while not necessarily maintaining band structure symmetry at various levels of development. The reader will find use of perturbational Feynman diagrammatic techniques, finite temperature quantum Matsubara Green's functions, and quantum perturbational derivations without the diagrams.

Superconductivity may be able to play a new or improved role in advancing technologies such as wireless communications, satellites, and other electronic and electromagnetically technologies which require the use of microscopic or nanoscopic materials, structures and devices. These all enlist small superconductors in 3D, 2D, 1D, and 0D, which includes bulk like systems, layered or thin film or atomic sheet systems [Osofsky et al., 2016a], [Osofsky et al. 2016b], nanowires/nanotubes/nanocables [Krowne, 2011], [Martinez et al., 2012], [Martinez et al., 2013], and quantum dots. Obtaining superconductors in such small systems necessitates relaxed requirements on critical values of temperature $T_c$, current $J_c$ and magnetic field $H_{c1}$ and $H_{c2}$ . One way to obtain larger values of all these critical constants, may be to introduce disorder intentionally.

We have found that functionalizing graphene [Osofsky et al., 2016a] with atoms of F, N, or O, allows one to dial into various matter states, which may be metallic, insulating, or even superconducting. This ability to dial in such characteristics at the nanoscopic or atomic level is something that was not attainable previously in a systematic fashion, which can be repeatedly duplicated now in the laboratory. This is the latest era of atomic and nanoscopic science: dimensions manipulated are on the order of a few angstroms. One of the remarkable facets of that study is despite the excitement generated by the achievement of metallic single layer graphene, it has been confused by the fact that seminal theoretical work [Abrahams et al., 1979] predicted that purely two-dimensional (2D) systems should not be metallic – because of disorder. The situation is confounded further by later theoretical work showing that Dirac Fermionic systems with no spin-orbit interactions and Gaussian correlated disorder exhibit scaling behavior, should always be metallic [Das Sarma et al., 2011]. So what is it, metallic or insulating? Really the answer is both due to the metal-insulator transition (MIT), which also suggests the possibility of dialing into a superconducting state.

Extremely thin metallic-like oxides, on the scale of ten nm's, such as $RuO_2$ thin film layers [Osofsky et al., 2016b], is a highly disordered conductor in which resistivity does not decrease with decreasing temperature, and the disorder-driven MIT is then described as a quantum phase transition characterized by extended states for the metallic phase and by localized states for the insulating phase. The existence of metallic behavior reported for this 2D material once again violates the famous prediction of Abrahams, Anderson, Licciardello, and Ramakrishnan [Abrahams et al., 1979] that all 2D systems must be localized regardless of the degree of disorder. The discovery of a metallic state in high-mobility metal oxide field-effect transistors (HMFET) motivated several theoretical approaches that included electron–electron interactions to screen disorder. These models adequately described the results for low carrier concentration, high-mobility systems, but are not applicable to the case of highly disordered 2D metals. And once again, the metal-insulator transition (MIT), suggests the possibility of dialing into a superconducting state.

Dimensions for $RuO_2$ nanoparticles which decorate $SiO_2$ inner core nanowires, forming nanocables [Krowne, 2011], are on the order of 3 nm. These 1D type of structures, might be investigated for possibilities of superconductivity. Although modeling makes some substantial simplifications, looking at the continuous cylindrical coverage of $RuO_2$ as a nanotube, in fact it is composed of somewhat randomly placed popcorns of $RuO_2$ nodules. Although early papers, for example, the one by Anderson [Anderson, 1959], have made some qualitative arguments for disorder effects based upon some first and second order quantum mechanical perturbation theory, the theoretical and experimental insights dependent upon atomic and nanoscopic engineering of materials, did not exist back then, and so limits the reach of such earlier works.

However, ultrasmall samples used in tunneling experiments, led to studies calling into question use of the grand canonical ensemble to describe electron pairing gap $\Delta_0(n_e)$, using an attractive Hubbard model (in real space) to represent s-wave superconductors, obtaining the gap parameter $\Delta_{N_e}(n_e, N)$ varying with averaged electron density $n_e$ and site number $N$ [Tanaka and Marsiglio, 2000a]. In [Tanaka and Marsiglio, 2000], and in [Tanaka and Marsiglio, 2000b], the suggested approach in [Anderson, 1959], solving for the eigenvalues and eigenstates of a non-interacting problem, diagonalizing the single-particle Hamiltonian, finding the transformed electron-electron interaction, and then applying the BCS variational procedure, generates a modified BCS gap equation. An alternative view using the effective Hamiltonian, diagonalizing by a Bogoliubov-Valatin transformation with a de Gennes approach, allows inspection of specific sites with one added impurity atom [Tanaka and Marsiglio, 2000b]. Such studies point up the importance of looking at the various nanoscopic aspects of superconductivity, although in the work here, we will not address such site by site nonuniformities (with either BCS and/or BdG tactic).

Superconductivity can be modified by various effects related to randomness, disorder, structural defects, and other similar physical effects. Their affects on superconductivity is important because such effects are intrinsic to certain material system's preparation, or may be intentionally produced. For example, in the 1980s on Anderson localization and dirty superconductors $H_{c2}$ is affected [G. Kotliar and A. Kapitulnik, 1986]; and high $T_c$ superconductors with structural disorder affects the electron-electron attraction stipulated by exchange of low-energy excitations, with substantial enhancement of $T_c$ [Maleyev and B. P. Toperverg, 1988]. In the 1990s, looking at thermal fluctuations, quenched disorder, phase transitions, and transport in type-II superconductors, vortices are pinned due to impurities or other defects which destroys long range correlations of the vortex lattice [D. S. Fisher, M. P. A. Fisher, D. A. Huse, 1991]; in field-induced superconductivity in disordered wire networks, small transverse magnetic applied fields increased the mean $T_c$ in disordered networks [Bonetto et al., 1998]; for structural disorder and its effect on the superconducting transition temperature in the organic superconductor κ-$(BEDT\text{-}TTF)_2Cu[N(CN)_2]Br$, $T_c$ is reduced in quenched cooled state [Su et al., 1998]; enhancement of $J_c$ density in single-crystal $Bi_2Sr_2CaCu_2O_8$ superconductors occurs by chemically induced disorder [Wang et al., 1990]. In the 2000s, surface enhancement of superconductivity occurred in single crystal tin, due to cold worked surface with surface enhanced order parameter [Kozhevnikov, 2005]; for disorder and quantum fluctuations in superconducting films in strong magnetic fields, $H_{c2}$ can increase and especially at low temperature [Galitski, 2001]; preparing amorphous $MgB_2/MgO$ superstructures which produces a model disordered superconductor, bilayers were made with relatively high $T_c$ [Siemons et al., 2008]; disorder-induced superconductivity can be produced in ropes of carbon nanotubes, with $T_c$ increasing with disorder [Bellafi, Haddad, and Charfi-Kaddour, 2009]; disordered 2D superconductors are examined for the role of temperature and interaction strength in the Hubbard model when the on-site attraction is switched off on a fraction $f$ of sites while keeping the attraction $U$ on the remaining sites, showing that near $f = 0.07$, $T_c$ increases with $U$ in the $2 \leq U \leq 6$ range [Mondaini, 2008]; enhancement of the high magnetic field $J_c$ of superconducting $MgB_2$ by proton irradiation occurs, with the irradiation pinning the vortices increasing $J_c$ [Bugoslavsky, 2001]; examination of the Lindemann criterion and vortex phase transitions in type-II superconductors, shows the destruction of vortex order by random point pinning and thermal fluctuations [Kierfeld and V. Vinokur, 2004]; doping induced disorder and superconductivity properties in carbohydrate doped $MgB_2$, increases the $J_c$ density [Kim et al., 2008]; the interplay between superconductivity and charge density waves is affected by disorder [Attanasi, 2008]; insensitivity of d-wave pairing to disorder in the high temperature cuprate superconductors, increase then decrease from scattering with weak dependence of $T_c$ on $n_{imp}$ impurity in theory as compared to experimental results [Kemper et al., 2009].

In the 2010s, for strongly disordered TiN and NbTiN s-wave superconductors probed by microwave electrodynamics [Driessen, 2012], it is mentioned that a decreased $T_c$ with increasing sheet resistance is found for MoGe films by Finkelstein [Finkelstein, 1987]; dynamical conductivity across the disorder tuned superconductor-insulator transition, has disorder enhanced absorption in conductivity and expands the quantum critical region [Swanson et al., 2014]; effects of randomness on $T_c$ in quasi-2D organic superconductors, leads to lowered $T_c$ [Nakhmedov, Alekperov, and Oppermann, 2012].

Here we show, in the context of a Cooper instability relationship, that introduction of disorder through impurities could possibly lead to an increase in $T_c$, using one of the simplest reductions from the hierarchy of complex representations. Other reductions in the hierarchy may show increases then decreases, depending on the specific reductions used. Although there is a complicated collection of quantum Green's functions, polarization functions (correlation functions), and vertex functions, used in the hierarchy of assumptions to extract out an analytical formula for $T_c$, one may obtain at least one formula from it which is recognizable, and relatable to BCS theory. Handling modified BCS superconductors, medium temperature superconductors, and high $T_c$ superconductors may be possible by appropriate use of the more accurate function representations and appropriate adaptions in the hierarchy.

Below in the following sections will be discussed the phonon operator utilized (Section II), the electron-phonon interaction potential, the Matsubara quantum many-body Green's functions employed for phonons and electrons (Section III), determination of the perturbed electron Matsubara quantum many-body Green's from its bare value, an examination of how the perturbed phonon quantum Green's function may be handled in a simplified manner and ramifications (Section IV), finding the renormalized phonon Green's function due to electron screening (Section V), electron vertex renormalization due to phonons (Section VI), renormalized total potential interaction energy due to Coulomb and phonon effects (Section VII), how the RPA permittivity is controlled through the charge-charge polarization diagram and related issues (Section VIII), disorder characterization for impurity scattering (Section IX), obtaining the ladder superconducting Cooper vertex (Section X) with the incorporation of impurity scattering, and a Cooper instability equation which is solvable in a hierarchy of possible equations of the critical temperature $T_c$ (Section XI). Section XII relates the gap parameter to the disorder potential energy, giving the indirect relationship between $T_c$ and gap parameter $\Delta$ when obtaining $\Delta$'s dependence on the same disorder potential energy quantity as $T_c$. Section XIII offers conclusions.

All of this analysis is done while not necessarily maintaining band structure symmetry at various levels of development. We note that the interested reader can refer to earlier works involving elementary excitations in solids, quasiparticles, and many-body effects in [Pines, 1964], [Pines, 1979] and [Ginzburg and Kirzhnits, 1982], and [Kittel, 1987] (which includes magnons). Those interested in a more refined electron-phonon interaction model (beyond the two-valued step function model) using the spectral density $\alpha^2(\omega)F(\omega)$ , can look at the very thorough review of [Corbette, 1990], which provides detailed information on the real and imaginary axis analytical continuation relation between, of the solution to the [Eliashberg, 1960a], [Eliashberg, 1960b] equations for the non-weak or strong coupling regimes. It is interesting that with some slight modifications, the strong case for the $T_c$ expression is similar to the weak form, which is treated herein. See [Bautolf, 2016] for a recent discussion of this and other topics in superconductivity, such as Ginzburg-Landau theory, the two fluid model of Gorter and Casimir, type-II superconductivity with vortices – including work by [Abrikosov, 2004], [Karnaukhov and Shepelev, 2008], and the London theory; [Tinkham, 1980], [Schrieffer, 1964] for details on the microscopic approach, as well as [Abrikovsov, Gorkov, and Dzyaloshinski, 1963], [Abrikovsov, Gorkov, and Dzyaloshinski, 1965] and [Abrikovsov and Gorkov, 1959a] (electrodynamics using a Matsubara thermodynamic approach for $\delta$, relationship between **j** and **A** [i.e., $Q$ hitting **A**], and penetration depth $\delta$), [Abrikovsov and Gorkov, 1959b] (obtains $T = 0$ electrodynamics equations, with introduction of the **A** photonic field Green's function $D$ and the system fermions Green's functions $G$ and $F$, again finding $Q$ and $\delta$), [Abrikovsov and Gorkov, 1961] (magnetic type impurities are expected to break the time reversal symmetry of the Cooper pairs, so although this is interesting work, our intent in here is to use materials which do not intentionally have such symmetry breaking.). For a review of localized impurity states, see [Balatsky, Vekhter and Zhu, 2006].

## II. PHONON OPERATOR AND THE ELECTRON-PHONON INTERACTION

The complete quantum phonon operator in second quantization format $A$, requires a raising $b^{\dagger}$ and lowering operator $b$. Denoting the reciprocal space phonon momentum as **q**, with **G** being the reciprocal lattice Umklapp vector, the $A$ operator is simply the sum of the two second quantization operators, with appropriate momentum indices.

$$A_{\mathbf{q}+\mathbf{G},\lambda_b,\lambda_p} = b^{\dagger}_{-(\mathbf{q}+\mathbf{G}),\lambda_b,\lambda_p} + b_{\mathbf{q}+\mathbf{G},\lambda_b,\lambda_p} \qquad (1)$$

Here $\lambda_b$ and $\lambda_p$ are respectively the phonon branch type (acoustic or optical) and phonon polarization type (longitudinal or transverse [two possibilities]). $b^{\dagger}$ and $b$ exist in the phonon Hilbert space. In general, the raising and lowering operators for a particular phonon quantum state depend on the surrounding sea of other phonons, so $b^{\dagger}$ and $b$ differ from their bare operators $b_0^{\dagger}$ and $b_0$, which exist without such influences or interactions. This point will become very important later on in deriving an acceptable form for the overall problem at hand.

Recognizing that the product space $|\psi_{el}\rangle \otimes |\psi_{ph}\rangle$ must be used for the electron-phonon interaction energy $V_{e\text{-}ph}$ ,

$$V_{e-ph} = \frac{1}{\mathcal{V}} \sum_{\mathbf{k}\in FBZ_e} \sum_{\sigma} \sum_{\mathbf{q}\in FBZ_p} \sum_{\lambda_b,\lambda_p} \sum_{\mathbf{G}\in RL} g_{\mathbf{q},\mathbf{G},\lambda_b,\lambda_p} c^{\dagger}_{\mathbf{k}+\mathbf{q}+\mathbf{G},\sigma} c_{\mathbf{k},\sigma} A_{\mathbf{q}+\mathbf{G},\lambda_b,\lambda_p} \qquad (2)$$

where $g_{\mathbf{q},\mathbf{G},\lambda_b,\lambda_p}$ is the degeneracy factor for the phonons, $c^{\dagger}$ and $c$ are the destruction and creation electron operators in their electron Hilbert space, and $\mathcal{V}$ is the real space volume. The summations occur respectively for **k**, $\sigma$, **q**, and **G** in the 1st Brillouin zone ($FBZ_e$) of the electron reciprocal lattice, the two electron spin polarizations, in the 1st Brillouin zone ($FBZ_{ph}$) for phonons, and in the reciprocal lattice.

## III. MATSUBARA MANY-BODY QUANTUM GREEN'S FUNCTIONS FOR PHONONS AND ELECTRONS

The imaginary time based Matsubara quantum many-body Green's functions can be convenient to use, and we do that here for both the phonon and electron propagators. First look at the phonon propagator,

$$D_\lambda(\mathbf{q}, \tau) = -\left\langle T_\tau \left[ A_{\mathbf{q},\lambda}(\tau) A_{-\mathbf{q},\lambda}(0) \right] \right\rangle_0 \qquad (3)$$

where the compressed notation form $\lambda = (\lambda_b, \lambda_p)$ is used, and the Umklapp process is reduced to within the 1st Brillouin zone, i.e., $\mathbf{G} = 0$. $T_\tau$ is the time ordering operator in the imaginary Matsubara time $\tau$ frame. Subscript "0" on the outside of the brackets indicates thermal averaging over the ensemble.

Next turn ones attention to the electron propagator, expressing it again in terms of the imaginary time $\tau$ ,

$$G_\sigma(\mathbf{k}, \tau) = -\frac{\left\langle T_\tau \left\{ U(\beta, 0) c_{\mathbf{k},\sigma}(\tau) c^\dagger_{\mathbf{k},\sigma}(0) \right\} \right\rangle_0}{\left\langle U(\beta, 0) \right\rangle_0} \qquad (4)$$

where $U(\beta, 0)$ is the unitary evolution operator in the imaginary Matsubara time frame, evaluated at $\tau = \beta$ . It depends on the time ordering of the perturbing interaction energies $P(\tau_i) = V_{e\text{-}ph}(\tau_i)$. Here $\beta = \hbar / (k_B T)$. The evolution operator for an arbitrary time $\tau$ is

$$U(\tau, 0) = \sum_{m=0}^{\infty} \frac{1}{m!} (-1)^m \int_o^\tau d\tau_1 \int_0^\tau d\tau_2 \cdots \int_0^\tau d\tau_m T_\tau \left\{ P(\tau_1) P(\tau_2) \cdots P(\tau_m) \right\} \qquad (5)$$

This formula has assumed the perturbation energy potential acts nearly instantaneously, which is why the $P(\tau_i)$'s are a single function of imaginary time $\tau_i$ ; i = 1, 2, 3, ⋯ $m$. But, in fact, for phonons as seen by two explicit similarly expressed forms for the pure electron-electron Coulomb interaction $W^{Coul}(\tau)$ and the electron-electron mediated phonon interaction $P(\tau_i, \tau_j) = V_{e\text{-}ph}(\tau_i, \tau_j)$ ,

$$V_{e-ph}(\tau_i, \tau_j) = \frac{1}{2\mathcal{V}} \sum_{\mathbf{k}_1,\sigma_1} \sum_{\mathbf{k}_2,\sigma_2} \sum_{\mathbf{q},\lambda} \frac{1}{\mathcal{V}} |g_{\mathbf{q},\lambda}|^2 D^0_\lambda(\mathbf{q}, \tau_i - \tau_j) c^\dagger_{\mathbf{k}_1+\mathbf{q},\sigma_1}(\tau_j) c^\dagger_{\mathbf{k}_2-\mathbf{q},\sigma_2}(\tau_i) c_{\mathbf{k}_2,\sigma_2}(\tau_i) c_{\mathbf{k}_1,\sigma_1}(\tau_j) \qquad (6)$$

$$W^{Coul}(\tau) = \frac{1}{2\mathcal{V}} \sum_{\mathbf{k}_1,\sigma_1} \sum_{\mathbf{k}_2,\sigma_2} \sum_{\mathbf{q}\neq 0} V_{\mathbf{q}} c^\dagger_{\mathbf{k}_1+\mathbf{q},\sigma_1}(\tau) c^\dagger_{\mathbf{k}_2-\mathbf{q},\sigma_2}(\tau) c_{\mathbf{k}_2,\sigma_2}(\tau) c_{\mathbf{k}_1,\sigma_1}(\tau) \qquad (7)$$

the phonon mediated interaction takes a finite time to occur compared to the pure Coulomb interaction. Thus, $\Delta\tau^{lt} = \tau_i^{lt} - \tau_j^{lt} << \Delta\tau^{ph} = \tau_i^{ph} - \tau_j^{ph}$ . Equation (7) has merely used the fact that photons mediate the Coulomb interaction, and in a nonrelativistic approximation, it is satisfactorily to set $\Delta\tau^{lt} \approx 0$ , or to assign one time to $\tau_i^{lt}$ and $\tau_j^{lt}$ .

Note that the two particle Coulomb interaction operator $W^{Coul}(\tau)$ containing four $c$ electron operators with Coulomb interaction potential energy $V_q$, has $V_q$ replaced by $|g_{q\lambda}|^2$ $D(q, \tau_i - \tau_j)$ for phonon mediated e-e interaction. The phonon mediated e-e interaction has used the bare (unperturbed) phonon Green's function $D^0_\lambda(\mathbf{q}, \tau_i - \tau_j)$ , whose justification will be covered next.

## IV. APPROXIMATION OF THE PHONON MATSUBARA PROPAGATOR BY ITS BARE VALUE

It is worthwhile to make an argument as to why it might be propitious to simplify the Green's function for the phonon from one dependent on the perturbed many-body form to one given by its bare unperturbed form. The guiding reasoning is this. If the complete and general Green's functions for both the electrons and phonons are used, phonon interactions will be renormalized by the many-body electron effects, but also the electron effects will be renormalized by the phonon many-body effects. This would constitute what we could refer to as simultaneous double renormalizations. Clearly, this would be the most accurate way to do the calculations. But we suspect, because the phonons originate from ionic motion, and ions are extremely massive compared to the electron mass, that the electron effect may dominate because of their much higher velocity. Of course, a somewhat counter argument is that it is really the momentum that is important, and that the $p_e = m_e v_e$ product may not exceed that of the phonon momentum $p_{ph}$, because $M_{ion} >> m_e$.

A way out of this conundrum, is to evaluate the effect of additional phonon lines on an electron 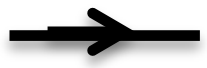 scatter off of a vertex 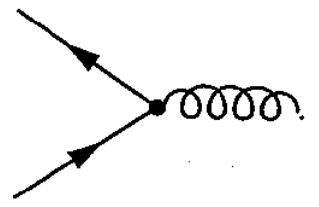 with a phonon 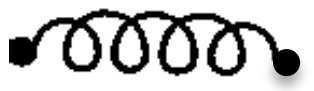, produced namely by

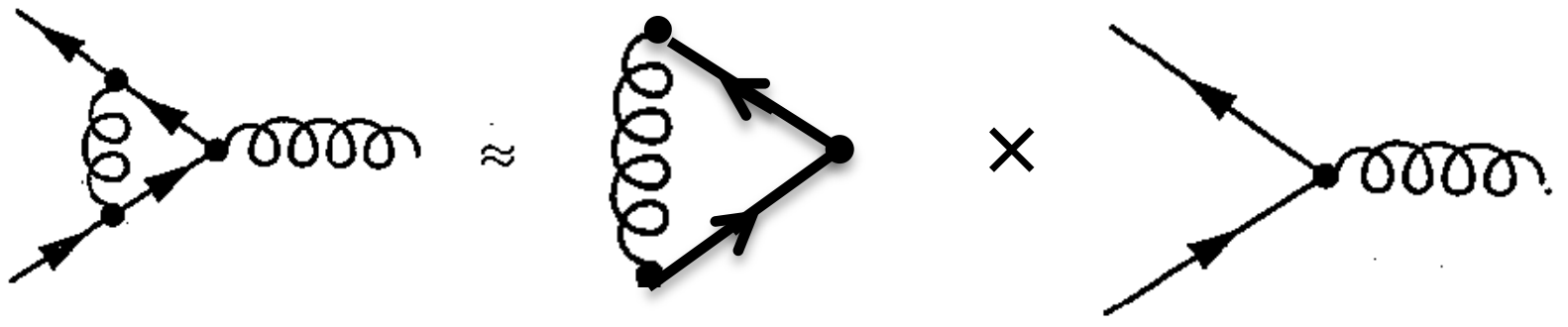

(8)

where the first product factor on the right hand side is about

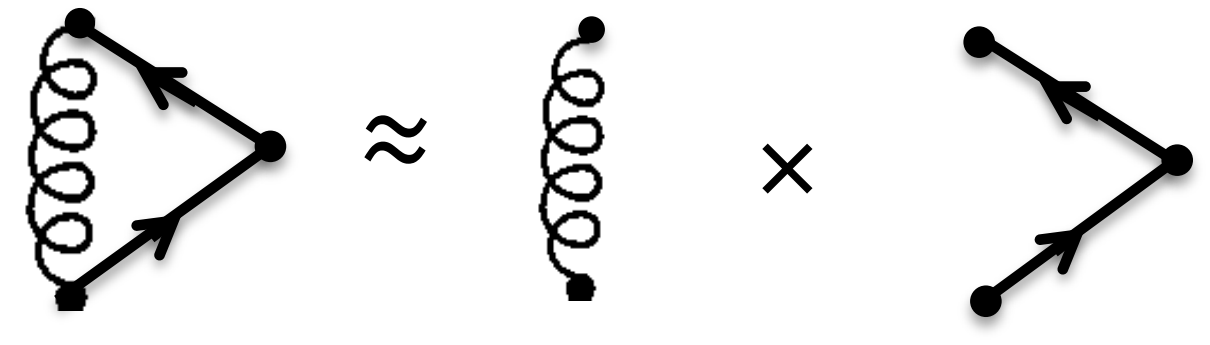

(9)

Close top and bottom terminations on the right most product on the right hand side of (9) as an approximation, giving us a pair bubble times the bare vertex:

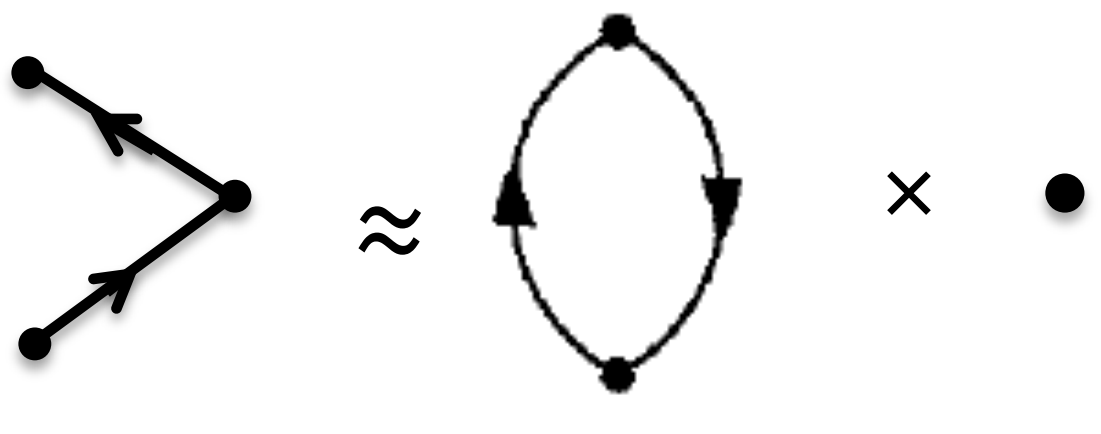

(10)

Here the polarization bubble [Fetter & Walecka, 1971] is

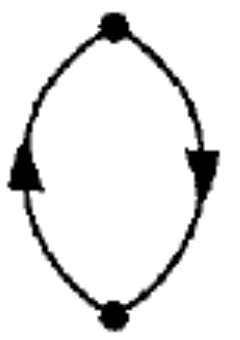

$$= \quad \chi_0(q, iq_n) = -\Pi_0(q, iq_n) = \chi_0^R(q, 0)\Big|_{\substack{low\ T \\ long\ \lambda_q}} = \chi_0(q, 0) \approx -d(\varepsilon_F)$$

(11)

Here $d(\varepsilon_F)$ is the density of states at the Fermi energy (units are in inverse $energy \cdot volume$). The bare phonon propagator $D_\lambda^0$, with its effective vertex strength $C_v^{eff}$, is

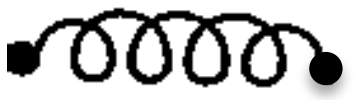

$$= \quad D_\lambda^0(\mathbf{q}, iq_n) = \frac{2\hbar\omega_q}{(iq_n)^2 - (\hbar\omega_q)^2}\Bigg|_{|iq_n| << \omega_q < \omega_D} \approx -\frac{2}{\hbar\omega_q} \tag{12}$$

and

$$C_v^{eff}(q) = \frac{1}{\mathcal{V}}|g_q|^2 \times \frac{\hbar\omega_D}{\varepsilon_F} = \frac{1}{2}W^{Coul}(q)\hbar\omega_q \times \frac{\hbar\omega_D}{\varepsilon_F} \approx \frac{1}{2}W^{R,\,Coul}(q)\Big|_{\substack{q\to 0, \\ iq_n = 0}}\hbar\omega_q \times \frac{\hbar\omega_D}{\varepsilon_F} \approx \frac{1}{2d(\varepsilon_F)}\hbar\omega_q \times \frac{\hbar\omega_D}{\varepsilon_F}$$

(13)

Result for the correction factor is (using Bohn-Staver expression for sound $v_s$)

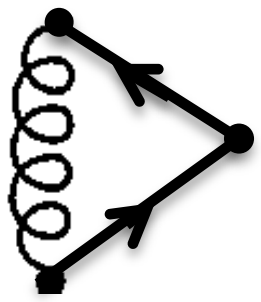

$$\approx \frac{\hbar\omega_D}{\varepsilon_F} = \frac{\hbar\omega_D}{p_F v_F/2} = \frac{\hbar\omega_D}{\hbar k_F v_F/2} = \frac{v_s k_D}{k_F v_F/2} = \sqrt{\frac{Zm_e}{3M_{atom}}}\frac{k_D}{k_F v_F/2} = 2\sqrt{\frac{Z}{3}}\sqrt{\frac{m_e}{M_{atom}}}\frac{k_D}{k_F} \approx \sqrt{\frac{m_e}{M_{atom}}}$$

(14)

giving for the correction term,

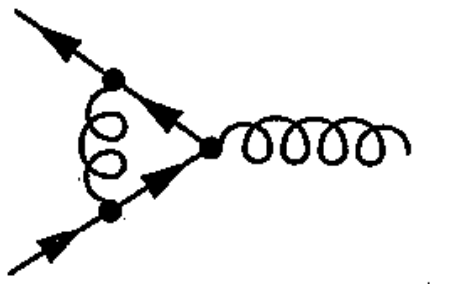

$$\approx \sqrt{\frac{m}{M}} \times$$

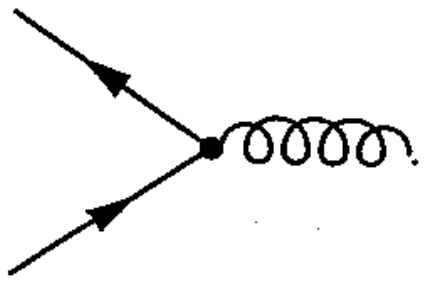

(15)

and showing, because $m_e/M_{ion} << 1$, that the renormalization process, may be considerably simplified, by dropping terms with vertices dressed by additional phonon lines. Such dropping of those terms is known as Migdal's theorem, and we enlist it to approximate the fully dressed (perturbed) phonon propagator $D_\lambda$, with the bare (unperturbed) propagator $D_\lambda^0$.

## V. FINDING THE RENORMALIZED PHONON GREEN'S FUNCTION DUE TO ELECTRON SCREENING

The phonon propagator may be renormalized in the RPA approximation, denoted by $D_\lambda^{RPA,\,ren}(\mathbf{q}, iq_n)$ or by $D_\lambda^{RPA}(\mathbf{q}, iq_n)$ for short. Diagrammatically, it is expressible in a Dyson equation form, that is it is equal to the bare phonon value plus a correction due to the RPA effect of the electron screening

$-D_{\lambda}^{RPA}(\mathbf{q}, iq_n) =$ 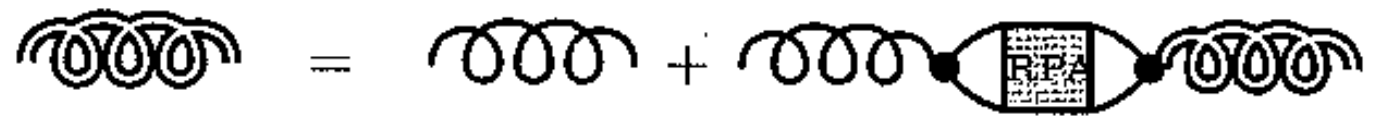 (16)

This Dyson like equation may be solved for the renormalized phonon Green's function:

$-D_{\lambda}^{RPA}(\mathbf{q}, iq_n) =$ 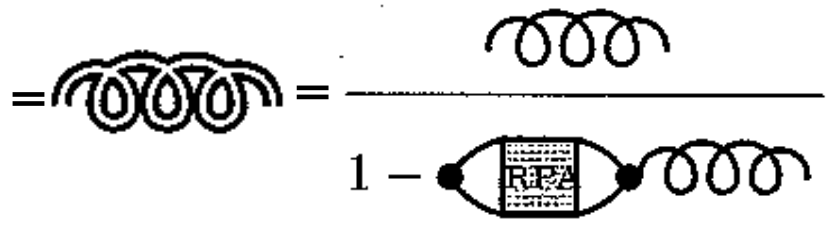 (17)

Phonon vertices •, from left to right, attached to the $-\chi^{RPA}(\tilde{q})$ electron screening RPA bubble 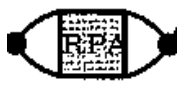, are given by respectively, $g_{\mathbf{q}\lambda}/\sqrt{\mathcal{V}}$ and $g_{\mathbf{q}\lambda}^{*}/\sqrt{\mathcal{V}}$. Equation (17) can be rewritten as

$$D_{\lambda}^{RPA}(\mathbf{q}, iq_n) = \frac{D_{\lambda}^{0}(\mathbf{q}, iq_n)}{1 - \frac{1}{\mathcal{V}}\left|g_{\mathbf{q}\lambda}\right|^2 D_{\lambda}^{0}(\mathbf{q}, iq_n)\chi^{RPA}(\mathbf{q}, iq_n)} \quad (18)$$

This expression may be evaluated with the bare phonon Green's function $D_{\lambda}^{0}(\mathbf{q}, iq_n)$,

$$D_{\lambda}^{0}(\mathbf{q}, iq_n) = \frac{2\Omega_{\mathbf{q}\lambda}}{\left(iq_n\right)^2 - \left(\Omega_{\mathbf{q}\lambda}\right)^2} \quad (19)$$

and the vertex products under the Jellium model,

$$\frac{1}{\mathcal{V}}\left|g_{\mathbf{q}\lambda}\right|^2 = \left(\frac{1}{\sqrt{\mathcal{V}}} g_{\mathbf{q}\lambda}\right)\left(\frac{1}{\sqrt{\mathcal{V}}} g_{\mathbf{q}\lambda}\right) = \frac{1}{2} W^{Coul}(q)\Omega_{\mathbf{q}\lambda} \quad (20)$$

Renormalized phonon propagator is then,

$$D_{\lambda}^{RPA,\, ren}(\mathbf{q}, iq_n) = \frac{2\Omega_{\mathbf{q}\lambda}}{\left(iq_n\right)^2 - \left(\Omega_{\mathbf{q}\lambda}\right)^2\left[1 + W^{Coul}(q)\chi^{RPA}(\mathbf{q}, iq_n)\right]} \quad (21)$$

Square bracketed expression is the inverse of the 4-reciprocal space RPA renormalized permittivity,

$$\varepsilon^{RPA}(\mathbf{q}, iq_n) = \frac{1}{1 + W^{Coul}(q)\chi^{RPA}(\mathbf{q}, iq_n)} \quad (22)$$

Final expression for renormalized phonon propagator is then,

$$D_{\lambda}^{RPA,\, ren}(\mathbf{q}, iq_n) = \frac{2\Omega_{\mathbf{q}\lambda}}{\left(iq_n\right)^2 - \left(\Omega_{\mathbf{q}\lambda}\right)^2/\varepsilon^{RPA}(\mathbf{q}, iq_n)} = \frac{2\Omega_{\mathbf{q}\lambda}}{\left(iq_n\right)^2 - \left(\omega_{\mathbf{q}\lambda}^{ren}\right)^2} \quad (23)$$

with the renormalized phonon frequency given by

$$\omega_{\mathbf{q}\lambda}^{ren} = \frac{\Omega_{\mathbf{q}\lambda}}{\sqrt{\varepsilon^{RPA}(\mathbf{q}, iq_n)}} \quad (24)$$

In the Jellium model, $\Omega_{\mathbf{q}\lambda}$ is actually independent of both **q** and λ, given by

$$\Omega_{\mathbf{q}\lambda}\Big|_{Jellium} = \Omega = \sqrt{\frac{Z^2 e^2 N_{ion}}{\varepsilon_0 M_{ion}\mathcal{V}}} = \sqrt{\frac{Z^2 e^2\left(\rho_{ion}^{0}\mathcal{V}\right)}{\varepsilon_0 M_{ion}\mathcal{V}}} = \sqrt{\frac{Z^2 e^2 \rho_{ion}^{0}}{\varepsilon_0 M_{ion}}} = \sqrt{\frac{Ze^2\left(Z\rho_{ion}^{0}\right)}{\varepsilon_0 M_{ion}}} = \sqrt{\frac{Ze^2 \rho_{el}^{0}}{\varepsilon_0 M_{ion}}} \quad (25)$$

## VI. RENORMALIZED ELECTRON VERTEX BASED ON PHONON MODIFICATION

We denote the modified bare vertex •, by 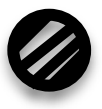, and see that it is expressed as

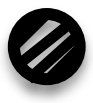 $= \frac{1}{\sqrt{\mathcal{V}}} g_{\mathbf{q}\lambda}^{RPA,\, ren} = \Big[ 1 \;+$ 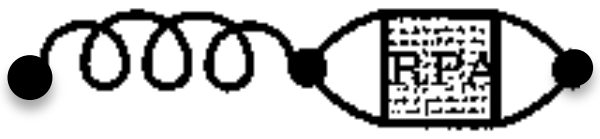 $\Big] \frac{1}{\sqrt{\mathcal{V}}} g_{\mathbf{q}\lambda}$ (26)

Inserting the expression for the phonon line and the RPA bubble 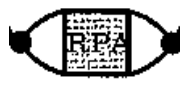 into (26),

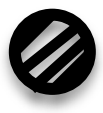 $= \frac{1}{\sqrt{\mathcal{V}}} g_{\mathbf{q}\lambda}^{RPA,\, ren} = \Big[ 1 \;+\; \{-W(q)\}\{-\chi^{RPA}(\tilde{q})\} \Big] \frac{1}{\sqrt{\mathcal{V}}} g_{\mathbf{q}\lambda}$ (27)

And stripping away the extra $1/\sqrt{\mathcal{V}}$ factors, we finally find the phonon mediated electron-electron vertex,

$$\begin{aligned} g_{\mathbf{q}\lambda}^{RPA,\, ren} &= \Big[ 1 \;+\; W(q)\chi^{RPA}(\tilde{q}) \Big] g_{\mathbf{q}\lambda} \\ &= \frac{g_{\mathbf{q}\lambda}}{\varepsilon^{RPA}(\mathbf{q}, iq_n)} \end{aligned} \tag{28}$$

which shows a renormalization by the permittivity.

## VII. RENORMALIZED TOTAL POTENTIAL ENERGY DUE TO COULOMB AND PHONON EFFECTS

There exists an effective potential energy $V_{eff}^{RPA,\, ren}$ which combines the effect of the repulsive Coulomb energy, RPA renormalized, and the attractive electron-electron phonon mediated energy, RPA renormalized,

$$V_{eff}^{RPA,\, ren} = W^{Coul,\, RPA,\, ren}(\mathbf{q}, iq_n) + \frac{1}{\mathcal{V}} \left| g_{\mathbf{q}\lambda}^{RPA,\, ren} \right|^2 D_{\lambda}^{RPA,\, ren}(\mathbf{q}, iq_n) \tag{29}$$

Diagrammatically, it looks like

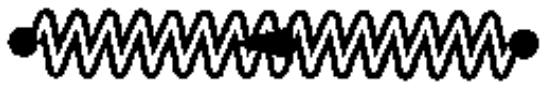 = 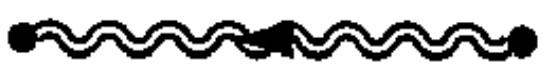 + 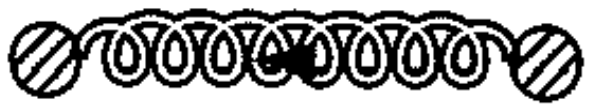 (30)

where the renormalized phonon propagator with its vertices is

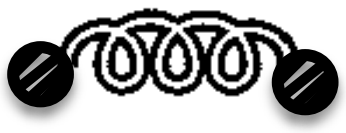 $= -\frac{1}{\mathcal{V}} \left| g_{\mathbf{q}\lambda} \right|^2 D_{\lambda}^{RPA}(\mathbf{q}, iq_n)$

$$= -\frac{W^{Coul}(q)}{\varepsilon^{RPA}(\tilde{q})} \frac{\left(\omega_{\mathbf{q}\lambda}^{ren}\right)^2}{\left(iq_n\right)^2 - \left(\omega_{\mathbf{q}\lambda}^{ren}\right)^2} = -\, W^{Coul,\, RPA,\, ren}(\tilde{q}) \frac{\left(\omega_{\mathbf{q}\lambda}^{ren}\right)^2}{\left(iq_n\right)^2 - \left(\omega_{\mathbf{q}\lambda}^{ren}\right)^2} \tag{31}$$

and where the renormalized Coulomb interaction is identified as

$$W^{Coul,\, RPA,\, ren}(\tilde{q}) = \frac{W^{Coul}(q)}{\varepsilon^{RPA}(\tilde{q})} \tag{32}$$

Reduction of the bare Coulomb effect seen in (32), is like that in earlier works [Ginzburg & Kirzhnits, 1982], (1.28) and (4.1) giving, respectively, $U_c^* = U_c\left[1 + U_c \ln\left(\omega_F / \omega_c\right)\right]^{-1}$, $\mu^* = \mu\left[1 + U_c \ln\left(E_F / \bar{\omega}\right)\right]^{-1}$; change $\bar{\omega}$ to $\omega$ in [Kittel, 1987] for (2.36).

## VIII. RPA PERMITTIVITY AS AFFECTED BY THE CORRELATION FUNCTION POLARIZATION

RPA permittivity is controlled through the correlation function, the charge-charge polarization $\chi$. Its expression in reciprocal 4-space is [Fetter and Walecka, 1971]

$$\varepsilon^{RPA}(\mathbf{q}, iq_n) = 1 - \frac{e^2}{\varepsilon_0 q^2}\chi_0(\mathbf{q}, iq_n) \tag{33}$$

What happens at the low frequency limit? By analytic continuation, the retarded polarization $\chi^R$ at any frequency $\omega$ is

$$\chi_0^R(\mathbf{q}, \omega) = \chi_0(\mathbf{q}, iq_n \to \omega + i\eta) \quad ; \quad \eta = 0^+ \tag{34}$$

However, it is known that as $\omega \to 0$ that

$$\chi_0^R(\mathbf{q}, \omega)\Big|_{\omega\to 0} \approx -d(\varepsilon_F) \tag{35}$$

where $d$ is the density of states (DOS) at the Fermi level $\varepsilon_F$, given in $energy^{-1} \cdot volume^{-1}$ units. But in the same limit,

$$\chi_0^R(\mathbf{q}, \omega)\Big|_{\omega\to 0} = \chi_0(\mathbf{q}, 0) \tag{36}$$

which makes

$$\chi_0(\mathbf{q}, 0) \approx -d(\varepsilon_F) \tag{37}$$

So the polarization we need in (33) is determined by the density of states at the Fermi level $\varepsilon_F$, which in 3D systems looks like

$$d^{3D}(\varepsilon_F) = \frac{1}{2\pi^2}\left(\frac{2m_e}{\hbar^2}\right)^{3/2}\left(\varepsilon_F\right)^{1/2} = \frac{m_e k_F}{\pi^2\hbar^2} \quad ; \quad \varepsilon_F = \frac{\hbar^2 (k_F)^2}{2m_e} \tag{38}$$

making the Fermi-Thomas screening wavenumber, squared,

$$\left(k_s^{3D}\right)^2 = -4\pi\left(e_0\right)^2 \chi_0^{3D}(0, 0) = \frac{4}{\pi}\frac{k_F}{a_0} \quad ; \quad e_0 = \frac{\hbar^2}{m_e a_0} \tag{39}$$

RPA permittivity is then

$$\varepsilon^{RPA,\,3D}(\mathbf{q}, iq_n)\Big|_{iq_n\to 0} \approx 1 + \frac{(k_s^{3D})^2}{q^2} = 1 + \frac{4k_F}{\pi a_0}\frac{1}{q^2} \tag{40}$$

In the small wavelength limit $q = 2\pi/\lambda$, $q$ becomes very large, and $\varepsilon^{RPA,\,3D}(\mathbf{q}, iq_n)\Big|_{iq_n\to 0} \to 1$, or permittivity goes to unity. In the opposite limit for large wavelengths, $q$ approaches zero, and

$$\varepsilon^{RPA,\,3D}(\mathbf{q}, iq_n)\Big|_{\substack{\mathbf{q}\to 0 \\ iq_n\to 0}} \approx \frac{(k_s^{3D})^2}{q^2} = \frac{4k_F}{\pi a_0}\frac{1}{q^2} \tag{41}$$

This makes renormalized phonon frequency

$$\omega_{\mathbf{q}\lambda}^{ren} = \sqrt{\frac{Ze^2\rho_{el}^0}{\varepsilon^{RPA}(\mathbf{q}, iq_n)\varepsilon_0 M_{ion}}} \tag{42}$$

become, using the formula for electron density

$$\rho_{el}^0 = n^{3D} = \frac{(k_F)^3}{3\pi^2} \tag{43}$$

the low frequency large wavelength result, with a linear frequency vs. momentum $q$ relationship,

$$\omega_{\mathbf{q}\lambda}^{ren} = \sqrt{\frac{Zm_e}{3M_{ion}}}\, v_F q = v_{s,\,ac} q \quad ; \quad v_{s,\,ac} = \sqrt{\frac{Zm_e}{3M_{ion}}}\, v_F \tag{44}$$

This gives us the Bohm-Staver relationship, for the acoustic dispersion of the phonons, from the Jellium model. The original optical Jellium phonons, get screened, and reduced to low frequencies. This happens because the RPA permittivity is enormous, proportional to $1/q^2$, and approaches infinity as $q$ goes to 0.

For 2D ordinary metals, and atomic layered graphene, various changes have to occur, including in $\chi$ , to obtain the various quantities like $\varepsilon^{RPA,\,2D}(\mathbf{q}, iq_n)$ and $\omega_{\mathbf{q}\lambda}^{ren,\,2D}$. For 2D ordinary metals, the DOS is a constant,

$$\rho^{2D}(\varepsilon) = \frac{g_{s,\,b} m^*}{2\pi\hbar^2} \tag{45}$$

where $g_{s,b}$ is the degeneracy due to spin and bands, either conduction or valence bands. In contrast to ordinary 2D metals, graphene with a single atomic layer of carbon atoms, has an energy dependent DOS like 3D systems. However, the power dependence on energy converts from a square root to linear form, in a tight-binding model approach,

$$\rho^{graphene}(\varepsilon) = \frac{g_{s,\,c}\,|\,\varepsilon\,|}{2\pi\hbar^2 v_F^{gr}} \tag{46}$$

with $\varepsilon > 0$ for the top Dirac cone and $\varepsilon < 0$ for the bottom Dirac cone, and its Fermi velocity given by an overlap orbital nearest neighbor construction,

$$v_F^{gr} = -\frac{3t_{gr}^{nn} a_{C-C}}{2\hbar} = \frac{3\,|\,t_{gr}^{nn}\,|\,a_{C-C}}{2\hbar} = 9.7\times 10^7 \ cm/\sec \tag{47}$$

Here $g_{s,c}$ is the degeneracy due to spin and cones (6 for either the top or bottom Dirac cones), $a_{C\text{-}C}$ the nearest neighbor carbon-to-carbon atom spacing, and the energy

$$\bar{\varepsilon}_{\mathbf{q}}^{\lambda_c} = \lambda_c \hbar v_F q \quad ; \quad q = |\mathbf{q}| \quad ; \quad \lambda_c = \pm 1 \tag{48}$$

where $\lambda_c$ is the cone type, + 1 for upper, -1 for lower. This is the energy with respect to the single unique onsite energy of the C atoms, $\varepsilon_{onsite}$ , the quantity utilized in (46).

## IX. DISORDER CHARACTERIZED IMPURITY SCATTERING

Disorder can be described by scattering from charged impurity ions, but nonmagnetic, or spinless. The bare electron-ion interaction is represented by a single dashed line going between a vertex and a star symbolizing the ion (say the jth one here). The renormalized interaction is then expressed as

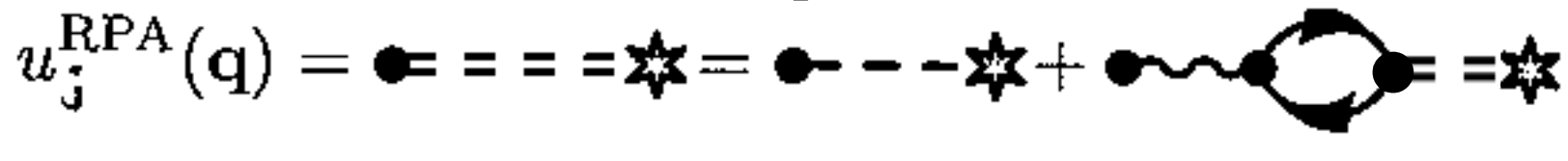

(49)

and written out as a equation,

$$u_j^{RPA}(\mathbf{q}) = u_j(\mathbf{q}) + \left\{-W^{Coul}(\mathbf{q})\right\}\left\{-\chi_0(\mathbf{q})\right\}u_j^{RPA}(\mathbf{q}) \tag{50}$$

This Dyson form of equation, can be readily solved, yielding

$$\begin{aligned} u_j^{RPA}(\mathbf{q}) &= \frac{u_j(\mathbf{q})}{1 - W^{Coul}(\mathbf{q})\chi_0(\mathbf{q})} \\ &= \frac{u_j(\mathbf{q})}{\varepsilon_0^{RPA}(\mathbf{q})} \end{aligned} \tag{51}$$

using a single bubble in the RPA expansion for $\chi$. Here the potential at the jth ion is found by lattice displacements $\boldsymbol{P}_j$ ,

$$u_j(\mathbf{q}) = u(\mathbf{q})e^{-i\mathbf{q}\cdot\mathbf{P}_j} \tag{52}$$

It is helpful to clarify at this point precisely what is meant by the RPA or random phase approximation. What it is a solution for the general sum of all diagrams for the correlation function,

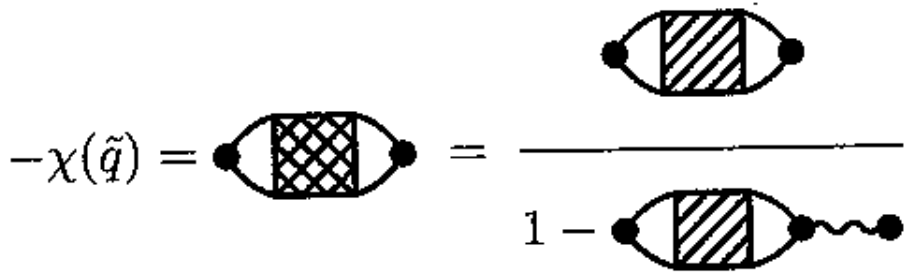

(53)

but with the irreducible diagrams correlation function $\chi^{irr}(\tilde{q})$ replaced in this Dyson solution of $\chi(\tilde{q})$, with a simple pair-bubble, or

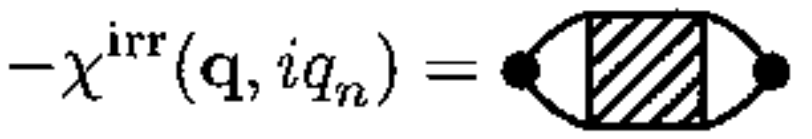

(54)

replaced by

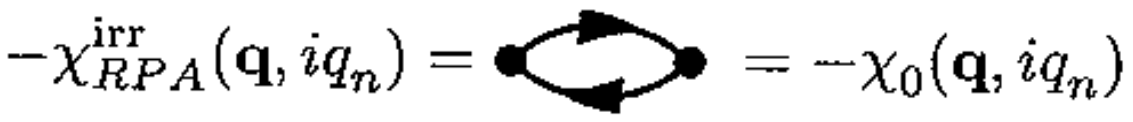

(55)

Solution of (52) depends not only on the simple polarization diagram, but also on $u_j(\mathbf{q})$ . For similar impurities all $u_j(\mathbf{q})$ would be the same, perhaps looking like in real space

$$u(\mathbf{r}) = -\frac{e_0^2}{|\mathbf{r}|}e^{-|\mathbf{r}|/a_d} \tag{56}$$

where $a_d$ is the characteristic decay distance. Its Fourier transform $u(\mathbf{q})$ is then

$$u(\mathbf{q}) = \iiint d^3r u(\mathbf{r})e^{-i\mathbf{q}\cdot\mathbf{r}} = -\frac{1}{4\pi\varepsilon_0}\frac{e^2}{\left(1/a_d\right)^2 + q^2} \tag{57}$$

## X. LADDER SUPERCONDUCTING COOPER VERTEX WITH DISORDER INCORPORATED

Superconducting vertex $\Lambda$ is a ladder sum of the total effective interaction 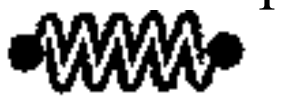 , which includes the renormalized electron-electron, mediated phonon, and impurity scattering effects,

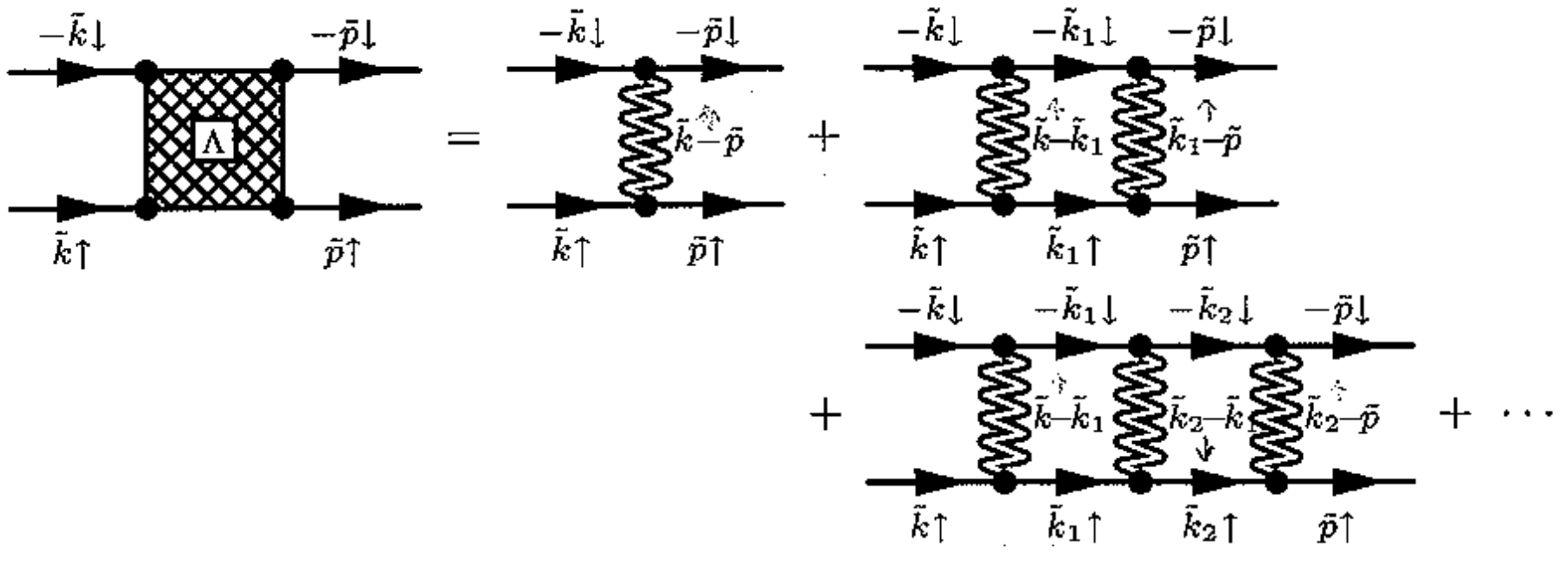

(58)

where $V^{RPA}_{eff,\,tot}$ is the total interaction due to these effects. Equation (58) can be restated as a Dyson type equation

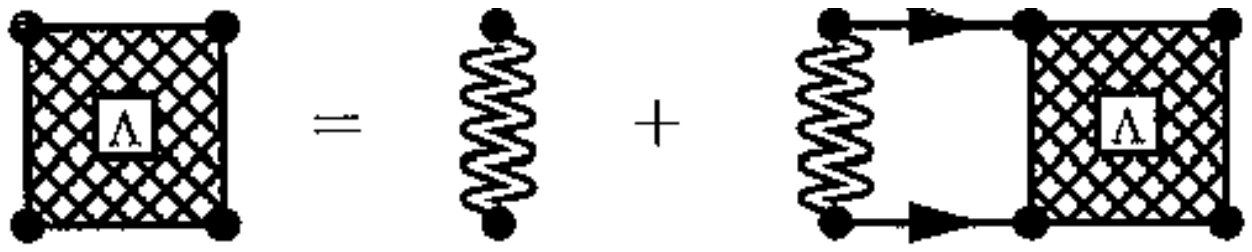

(59)

This vertex Dyson type equation, may be written out explicitly in terms of its vertex functions, quantum Green's functions, and effective total potential energy $V^{RPA}_{eff,\,tot}$ :

$$\Lambda(\tilde{k},\,\tilde{p}) = -\,V^{RPA}_{eff\ tot}(\tilde{k} - \tilde{p}) \ + \ \frac{1}{\mathcal{V}\beta}\sum_{\tilde{q},\,\lambda_{ph}}\left[-\,V^{RPA}_{eff\ tot}(\tilde{k} - \tilde{q})\right]G^0_{\uparrow}(\tilde{q})G^0_{\downarrow}(-\tilde{q})\Lambda(\tilde{q},\,\tilde{p}) \qquad (60)$$

Noting that $V^{RPA}_{eff,\,tot} < 0$, writing it as $V' = -\,V^{RPA}_{eff,\,tot}$, and dropping the super- and sub-scripts for brevity,

$$\Lambda(\tilde{k},\,\tilde{p}) = V'(\tilde{k} - \tilde{p}) \ + \ \frac{1}{\mathcal{V}\beta}\sum_{\tilde{q},\,\lambda_{ph}}\left[V'(\tilde{k} - \tilde{q})\right]G^0_{\uparrow}(\tilde{q})G^0_{\downarrow}(-\tilde{q})\Lambda(\tilde{q},\,\tilde{p}) \qquad (61)$$

Formally, this may be solved as follows:

$$\Lambda(\tilde{k},\,\tilde{p}) = \frac{V'(\tilde{k} - \tilde{p})}{\left\{1 \ - \ \dfrac{1}{\mathcal{V}\beta}\displaystyle\sum_{\tilde{q},\,\lambda_{ph}}\left[V'(\tilde{k} - \tilde{q})\right]G^0_{\uparrow}(\tilde{q})G^0_{\downarrow}(-\tilde{q})\dfrac{\Lambda(\tilde{q},\,\tilde{p})}{\Lambda(\tilde{k},\,\tilde{p})}\right\}} \qquad (62)$$

Equation (62) demonstrates the instability behavior when the 2nd denominator term approaches unity from the lower or upper sides.

## XI. CRITICAL TEMPERATURE OBTAINED FROM A COOPER INSTABILITY EQUATION

Let us examine the Cooper instability equation, obtained from equation (62) by setting the denominator to unity,

$$1 \ - \ \frac{1}{\mathcal{V}\beta}\sum_{\tilde{q},\,\lambda_{ph}}\left[V'(\tilde{k} - \tilde{q})\right]G^0_{\uparrow}(\tilde{q})G^0_{\downarrow}(-\tilde{q})\frac{\Lambda(\tilde{q},\,\tilde{p})}{\Lambda(\tilde{k},\,\tilde{p})} = 0 \qquad (63)$$

which should be evaluated at $T = T_c$, the critical temperature, since we expect the major physical change to occur here, and should be associated with the singularity. For the vertex function depending approximately only on its second variable, here $\tilde{p}$ , the instability condition becomes

$$1 - \frac{1}{\mathcal{V}\beta_c} \sum_{\tilde{q},\, \lambda_{ph}} \left[ V'(\tilde{k} - \tilde{q}) \right] G_{\uparrow}^{0}(\tilde{q}) G_{\downarrow}^{0}(-\tilde{q}) = 0 \tag{64}$$

which may be rewritten as

$$1 = \frac{1}{\mathcal{V}\beta_c} \sum_{\substack{iq_n \\ |iq_n| < \alpha\omega_D}} \sum_{\mathbf{q}} \sum_{\lambda_{ph}} \left[ V'(\tilde{k} - \tilde{q}) \right] G_{\uparrow}^{0}(\tilde{q}) G_{\downarrow}^{0}(-\tilde{q}) \tag{65}$$

Here $q_n$ is the Fermion index used in Matsubara reciprocal space imaginary frequency summations,

$$q_n = \frac{(2n+1)\pi}{\beta_c} = \frac{2\pi}{\beta_c}\left(n + \frac{1}{2}\right) \quad ; \quad \beta_c = \frac{1}{k_B T_c} \tag{66}$$

Bare Green's functions are

$$G_{\uparrow}^{0}(\mathbf{q}, iq_n) = \frac{1}{iq_n - \xi_{\mathbf{q}\uparrow}} \quad ; \quad G_{\downarrow}^{0}(\mathbf{q}, iq_n) = \frac{1}{iq_n - \xi_{\mathbf{q}\downarrow}} \tag{67}$$

Considering the case with no static or otherwise background magnetic field,

$$\xi_{\mathbf{q}\uparrow} = \xi_{\mathbf{q}\downarrow} = \xi_{\mathbf{q}} \tag{68}$$

Placing (67) and (68) into (65) then gives,

$$1 = \frac{1}{\mathcal{V}\beta_c} \sum_{\substack{iq_n \\ |iq_n| < \alpha\omega_D}} \sum_{\mathbf{q}} \sum_{\lambda_{ph}} \left[ V'(\tilde{k} - \tilde{q}) \right] \frac{1}{iq_n - \xi_{\mathbf{q}}} \frac{1}{-iq_n - \xi_{\mathbf{q}}} \tag{69}$$

The disorder effect contributes an extra diagram 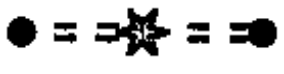 for renormalized impurity scattering to the effective potential energy due to the Coulomb scattering offsetting the attractive electron-electron phonon mediated scattering:

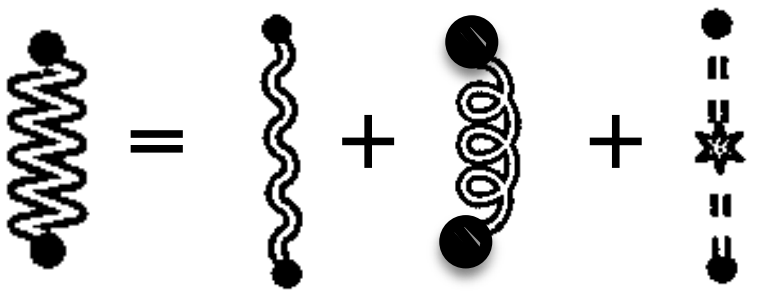

(70)

or

$$V_{eff}^{RPA,\, ren} = W^{Coul,\, RPA\,,\, ren}(\mathbf{q}, iq_n) + \frac{1}{\mathcal{V}} \left| g_{\mathbf{q}\lambda}^{RPA,\, ren} \right|^2 D_{\lambda}^{RPA,\, ren}(\mathbf{q}, iq_n) + W^{imp\ 1BA,\, RPA}(\mathbf{q}, iq_n) \tag{71}$$

where the last term is given in the 1st Born approximation when only scattering off of a single impurity scatter is allowed (Born approximation), and it is limited to one scattering event (1st). Under these assumptions, valid when scattering due to impurities is not too severe, the last term for impurity potential interaction energy is [Bruus and K. Flensberg, 2004]

$$W^{imp\ 1BA,\, RPA}(\mathbf{q}, iq_n) = n_{imp} \frac{u(\mathbf{q})}{\varepsilon^{RPA}(\mathbf{q}, 0)} \frac{u(-\mathbf{q})}{\varepsilon^{RPA}(-\mathbf{q}, 0)} \delta_{q_n,\, 0} \tag{72}$$

which we see has the renormalized impurity interaction lines for incoming $\mathbf{q}$ and outgoing $-\mathbf{q}$ momentum into the impurity center vertex, weighted by the impurity spatial density $n_{imp}$, which arises from the summation over all scatters per unit volume. The Kronecker delta function, forces one to consider Matsubara frequencies approaching zero.

In a hierarchy of approximations, with various physical effect ramifications, one simplification is to pull $V'(\tilde{k} - \tilde{q})$ out from under the $\mathbf{q}$ and $\lambda_{ph}$ summation signs, replacing it by an average. This process is consistent with the earlier two variable to one variable vertex function simplification, which suggests here we drop dependence on $\mathbf{k}$.

$$1 = \frac{V'_{av}}{\mathcal{V}\beta_c} \sum_{\substack{iq_n \\ |iq_n| < \alpha\omega_D}} \sum_{\mathbf{q}} \frac{1}{iq_n - \xi_{\mathbf{q}}} \frac{1}{-iq_n - \xi_{\mathbf{q}}} \tag{73}$$

Here the $\lambda_{ph}$ summation has been dropped entirely (and absorbed into $V'_{av}$) since the remaining bare Green's function contain no phonon polarization information. The $\mathbf{q}$ summation is implicitly over one spin type, where $\xi_q$ is taken above ($\xi_{\mathbf{q}} > 0$) and below ($\xi_{\mathbf{q}} < 0$) the Fermi surface. Above the Fermi surface $\xi_{\mathbf{q}}$ is unbounded, but below it can go to, at least in a normal 3D metal, to $-\varepsilon_F$ . Therefore, using the density of states $d_{st}(\xi)$ , (73) becomes

$$\begin{aligned}
1 &= \frac{V'_{av}}{\mathcal{V}\beta_c} \sum_{\substack{iq_n \\ |iq_n| < \alpha\omega_D}} \sum_{\mathbf{q}} \frac{1}{iq_n - \xi_{\mathbf{q}}} \frac{1}{-iq_n - \xi_{\mathbf{q}}} \\
&= \frac{V'_{av}}{\mathcal{V}\beta_c} \sum_{\substack{iq_n \\ |iq_n| < \alpha\omega_D}} \frac{1}{\mathcal{V}} \sum_{\mathbf{q}} \frac{1}{iq_n - \xi_{\mathbf{q}}} \frac{1}{-iq_n - \xi_{\mathbf{q}}} \\
&= \frac{V'_{av}}{\beta_c} \sum_{\substack{iq_n \\ |iq_n| < \alpha\omega_D}} \frac{1}{2} \int_{-\varepsilon_F}^{\infty} d_{st}(\xi) d\xi \frac{1}{iq_n - \xi_{\mathbf{q}}} \frac{1}{-iq_n - \xi_{\mathbf{q}}} \\
&= \frac{V'_{av}}{\beta_c} \sum_{\substack{iq_n \\ |iq_n| < \alpha\omega_D}} \frac{1}{2} \int_{-\varepsilon_F}^{\infty} d_{st}(\xi) d\xi \frac{1}{(q_n)^2 + \xi^2} \\
&\approx \frac{V'_{av}}{\beta_c} \sum_{\substack{iq_n \\ |iq_n| < \alpha\omega_D}} d_{st}(\xi = 0;\, \varepsilon = \varepsilon_F) \frac{1}{2} \int_{-\varepsilon_F}^{\infty} d\xi \frac{1}{(q_n)^2 + \xi^2} \\
&\approx \frac{V'_{av}}{\beta_c} \sum_{\substack{iq_n \\ |iq_n| < \alpha\omega_D}} d_{st}(\xi = 0;\, \varepsilon = \varepsilon_F) \frac{1}{2} \int_{-\infty}^{\infty} d\xi \frac{1}{(q_n)^2 + \xi^2} \\
&= \frac{V'_{av}}{\beta_c} \sum_{\substack{iq_n \\ |iq_n| < \alpha\omega_D}} \frac{d_{st}(\varepsilon_F)}{2} \frac{1}{q_n} \tan^{-1}\left(\frac{\xi}{q_n}\right)\Bigg|_{-\infty}^{+\infty} \\
&= \frac{V'_{av}}{\beta_c} \sum_{\substack{iq_n \\ |iq_n| < \alpha\omega_D}} \frac{d_{st}(\varepsilon_F)}{2} \frac{2}{q_n} \tan^{-1}(\infty)\,\mathrm{sgn}(q_n) \\
&= \frac{V'_{av}}{\beta_c} \sum_{\substack{iq_n \\ |iq_n| < \alpha\omega_D}} \frac{d_{st}(\varepsilon_F)}{2} \frac{2}{|q_n|} \frac{\pi}{2}
\end{aligned} \tag{74}$$

The final result obtained from the last line of equation (74), is a double-sided sum over Matsubara frequencies. The idea is to convert this into a single sided sum, convert that into an integral, and recognize a familiar form which has an analytical reduction. From (74) we write

$$1 = \frac{V'_{av}}{2\beta_c} \sum_{\substack{iq_n \\ |iq_n| < \alpha\omega_D}} d_{st}(\varepsilon_F) \frac{\pi}{|q_n|} = \frac{V'_{av} d_{st}(\varepsilon_F)}{2\beta_c} \sum_{\substack{iq_n \\ |iq_n| < \alpha\omega_D}} \frac{\pi}{|q_n|} \tag{75}$$

[Note that the earlier DOS $d(\varepsilon_F)$, and $d_{st}(\varepsilon_F)$, are identical, with the additional subscript merely serving to emphasize the distinction from the differential operator inside the integral.] Carrying out the steps outlined, using the Fermion frequency index relation (66),

$$
\begin{aligned}
1 &= \frac{V'_{av} d_{st}(\varepsilon_F)}{2\beta_c} \sum_{\substack{iq_n \\ |iq_n|<\alpha\omega_D}} \frac{\pi}{|q_n|} \\
&= \frac{V'_{av} d_{st}(\varepsilon_F)}{2\beta_c} \sum_{\substack{iq_n \\ |iq_n|<\alpha\omega_D}} \frac{\pi}{\left|\frac{(2n+1)\pi}{\beta_c}\right|} \\
&= \frac{V'_{av} d_{st}(\varepsilon_F)}{2\beta_c} \sum_{\substack{iq_n \\ |iq_n|<\alpha\omega_D}} \frac{\beta_c}{2} \frac{1}{|n+1/2|} \\
&= \frac{V'_{av} d_{st}(\varepsilon_F)}{2\beta_c} \frac{\beta_c}{2} \sum_{\substack{iq_n \\ |iq_n|<\alpha\omega_D}} \frac{1}{|n+1/2|} \\
&= \frac{V'_{av} d_{st}(\varepsilon_F)}{4} \sum_{\substack{n=0,\pm 1,\pm 2,\cdots \\ |q_n|<\alpha\omega_D}} \frac{1}{|n+1/2|} \\
&= \frac{V'_{av} d_{st}(\varepsilon_F)}{4} \left\{ \sum_{\substack{n=0,1,2,\cdots \\ |q_n|<\alpha\omega_D}} \frac{1}{n+1/2} + \sum_{\substack{n=-1,-2,\cdots \\ |q_n|<\alpha\omega_D}} \frac{1}{|n+1/2|} \right\} \\
&= \frac{V'_{av} d_{st}(\varepsilon_F)}{4} \left\{ \sum_{n=0}^{\alpha\beta_c\omega_D/(2\pi)-1/2} \frac{1}{n+1/2} + \sum_{n=-1}^{-\alpha\beta_c\omega_D/(2\pi)-1/2} \frac{1}{|n+1/2|} \right\} \\
&= \frac{V'_{av} d_{st}(\varepsilon_F)}{4} 2 \sum_{n=0}^{\alpha\beta_c\omega_D/(2\pi)-1/2} \frac{1}{n+1/2}
\end{aligned}
\tag{76}
$$

So we have obtained a single-sided sum as desired,

$$
1 = \frac{V'_{av} d_{st}(\varepsilon_F)}{2} \sum_{n=0}^{\alpha\beta_c\omega_D/(2\pi)-1/2} \frac{1}{n+1/2} \tag{77}
$$

One way to handle this summation is to replace it by an integral approximation, taking $1 \to dn$, and perform a change of variable, so that

$$
\begin{aligned}
1 &\approx \frac{V'_{av} d_{st}(\varepsilon_F)}{2} \int_0^{\alpha\beta_c\omega_D/(2\pi)-1/2} \frac{dn}{n+1/2} \\
&= \frac{V'_{av} d_{st}(\varepsilon_F)}{2} \int_{1/2}^{\alpha\beta_c\omega_D/(2\pi)-1/2+1/2} \frac{dm}{m} \\
&= \frac{V'_{av} d_{st}(\varepsilon_F)}{2} \ln(m)\Big|_{1/2}^{\alpha\beta_c\omega_D/(2\pi)} \\
&= \frac{V'_{av} d_{st}(\varepsilon_F)}{2} \left[ \ln\left(\frac{\alpha\beta_c\omega_D}{2\pi}\right) - \ln\left(\frac{1}{2}\right) \right] \\
&= \frac{V'_{av} d_{st}(\varepsilon_F)}{2} \ln\left(\frac{\alpha\beta_c\omega_D}{\pi}\right)
\end{aligned}
\tag{78}
$$

or

$$1 \;=\; \frac{V'_{av} d_{st}(\varepsilon_F)}{2} \ln\left(\frac{\alpha\beta_c\omega_D}{\pi}\right) \tag{79}$$

Equation (79) may be solved using (71) broken into two pieces, instead of three, the electron-electron and the electron-phonon renormalization terms collected together, and the impurity disorder contribution:

$$V'^{RPA,\,CP}_{eff}(\mathbf{q}, iq_n) \;=\; W^{Coul,\,RPA\,,\,ren}(\mathbf{q}, iq_n) + \frac{1}{\mathcal{V}}\left|g^{RPA,\,ren}_{\mathbf{q}\lambda}\right|^2 D^{RPA,\,ren}_{\lambda}(\mathbf{q}, iq_n) \tag{80}$$

$$V'^{RPA,\,imp}_{eff} \;=\; W^{imp\;1BA,\,RPA}(\mathbf{q}, iq_n) \tag{81}$$

to make the total as

$$V'^{RPA,\,ren}_{eff} \;=\; V'^{RPA,\,CP}_{eff} \;+\; V'^{RPA}_{imp} \tag{82}$$

[Notice that $V'^{RPA,\,CP}_{eff}$ of (80) is just the $V'^{RPA,\,CP}_{eff}$ found in (29).] When this decomposition is done, the solution to (79) may be written as

$$T_c \;=\; T^{\alpha}_c\Big|_{\substack{BCS\\pre}}\left(T_c\Big|_{\substack{BCS\\\exp}}\right)^{\gamma_{rp}} \tag{83}$$

where the following definitions are used:

$$T^{\alpha}_c\Big|_{\substack{BCS\\pre}} = \frac{\alpha}{\pi}\frac{\hbar\omega_D}{k_B} \;;\; T_c\Big|_{\substack{BCS\\\exp}} \;=\; e^{-\frac{2}{V'^{RPA,\,CP}_{eff,\,av}}\frac{1}{d(\varepsilon_F)}} \;;\; \gamma_{rp} \;=\; \frac{1}{1 \;+\; V'^{RPA}_{imp,\,av} \,/\, V'^{RPA,\,CP}_{eff,\,av}} \tag{84}$$

Here BCS as a subscript serves to identify the BCS like behavior in the prefactor coefficient and the exponential terms, whereas $\gamma_{rp}$ is the exponential modification power. Subscripts "av" on the renormalized potential energies indicates the following averaging process:

$$V'^{RPA}_{eff,\,av} \;=\; \left\langle V'^{RPA}_{eff}(\mathbf{q}, iq_n; T)\Big|_{iq_n\to\omega+i\eta}\right\rangle_{av} \;=\; V'^{RPA}_{eff}(\mathbf{q}_{av}, \omega_{av}; T) \tag{85}$$

$$V'^{RPA,\,CP}_{eff,\,av} \;=\; \left\langle V'^{RPA,\,CP}_{eff}(\mathbf{q}, iq_n; T)\Big|_{iq_n\to\omega+i\eta}\right\rangle_{av} \;=\; V'^{RPA,\,CP}_{eff}(\mathbf{q}_{av}, \omega_{av}; T) \tag{86a}$$

$$V'^{RPA}_{imp,\,av} \;=\; \left\langle V'^{RPA}_{imp}(\mathbf{q}, iq_n; T)\Big|_{iq_n\to\omega+i\eta}\right\rangle_{av} \;=\; V'^{RPA}_{imp}(\mathbf{q}_{av}, \omega_{av}; T) \tag{86b}$$

where analytic continuation is used.

By (72), (81) and (86), the $\gamma_{rp}$ may be transformed into an expression that implicitly shows the impurity or disorder density concentration $n_{imp}$,

$$\gamma_{rp} \;=\; \frac{1}{1 \;+\; \tilde{R}^{V}_{s,\,imp;\,CP}(n_{imp})} \tag{87}$$

where $\tilde{R}^{V}_{s,\,imp;\,CP}$ is the impurity to electron-phonon renormalized potential energy ratio, and in general may be expressed as,

$$\tilde{R}^{V}_{s,\,imp;\,CP}(n_{imp}) = \tilde{R}^{V\,(0)}_{s,\,imp;\,CP}(0) \;+\; \tilde{R}^{V\,(1)}_{s,\,imp;\,CP}(0)n_{imp} \;+\; \tilde{R}^{V\,(2)}_{s,\,imp;\,CP}(0)\left\{n_{imp}\right\}^2 \;+\; \cdots \tag{88}$$

Numbered superscripts (i) indicate the i$^{th}$ derivative taken. Inserting (88) into (87) gives

$$\gamma_{rp} \;=\; \frac{1}{1 \;+\; n_{imp}\left[\tilde{R}^{V\,(0)}_{s,\,imp;\,CP}(0)\left\{1/n_{imp}\right\} + \tilde{R}^{V\,(1)}_{s,\,imp;\,CP}(0) + \tilde{R}^{V\,(2)}_{s,\,imp;\,CP}(0)n_{imp} \;+\; \cdots\right]} \tag{89}$$

by factoring out a power of $n_{imp}$, suggested by (72). Clearly, the first term within brackets contains an unphysical singularity, and requires

$$\tilde{R}^{V\,(0)}_{s,\,imp;\,CP}(0) = 0 \tag{90}$$

giving

$$\gamma_{rp} \approx \frac{1}{1 + n_{imp}\tilde{R}^{V\,(1)}_{s,\,imp;\,CP}(0)} = \frac{1}{1 + n_{imp}R^{V}_{s,\,imp;\,CP}} \tag{91}$$

The next term, $\tilde{R}^{V\,(1)}_{s,\,imp;\,CP}(0) = R^{V}_{s,\,imp;\,CP}$, is the impurity to electron-phonon renormalized potential energy ratio per impurity scatterer, and may or may not differ from zero, depending upon details of the nanoscopic preparation of the impurity constituents. Equation (91) may be most consistent with non-alloying process of material preparation. Examination of the form of (91) shows that if the ratio is positive, with the impurity concentration always positive, the power modification exponential will be less than unity, and applied to the $T_c\big|_{\substack{BCS\\ \exp}}$ factor, which is a unitless number less than one, since it scales down the temperature prefactor $T_c^{\alpha}\big|_{\substack{BCS\\ pre}}$ which is in units of Kelvin, the new effective BCS exponential will be increased in size. This will have the effect of increasing $T_c$. On the other hand, if the ratio is negative, the power modification exponential will be greater than unity, and when applied to the $T_c\big|_{\substack{BCS\\ \exp}}$ factor, which is a unitless number less than one, since it scales down the temperature prefactor $T_c^{\alpha}\big|_{\substack{BCS\\ pre}}$, the new effective BCS exponential will decreased in size. This will have the effect of decreasing $T_c$.

If the first two terms in the brackets of (89), are zero, as some interpretations seem to have favored or implied previously {see [A. A. Abrikovsov and Gorkov, 1961] which states "It is well known that the transition temperature of superconductors with nonmagnetic impurities remains practically unchanged in the region of low concentrations", which cites [Abrikovsov, and Gorkov, 1959b] treating alloys at $T = 0$, and [A. A. Abrikovsov and Gorkov, 1959a] treating alloys at $T \neq 0$, although the latter two references do not seem explicitly to prove that assertion; or [Tinkham, 1980] on page 263 states "when we considered Anderson's theory of dirty superconductors [Anderson, 1959], i.e., nonmagnetic alloys with mean free path $\ell < \xi_0$, we note that pairing of time-reversed degenerate states led to same $T_c$ and BCS density of states as for a pure superconductor."; where [Anderson, 1959] made qualitative arguments for disorder effects based upon some first and second order quantum mechanical perturbation theory}, creating a stronger null as $n_{imp} \rightarrow 0$, that is

$$\tilde{R}^{V\,(0)}_{s,\,imp;\,CP}(0) = 0 \quad ; \quad \tilde{R}^{V\,(1)}_{s,\,imp;\,CP}(0) = 0 \tag{92}$$

then

$$\gamma_{rp} \approx \frac{1}{1 + \left(n_{imp}\right)^2 \tilde{R}^{V\,(2)}_{s,\,imp;\,CP}(0)} \tag{93}$$

Equation (93) may be most consistent with alloy processes. Note that both $\gamma_{rp}$ forms (91) and (93) do satisfy the necessary condition

$$\lim_{n_{imp}\to 0}\gamma_{rp} = 1 \qquad (94)$$

In order to have a weaker march to unity in (94), we will address the case of (91) below.

There is another subtlety to interpreting (91), and that is, increasing the number of impurities may reduce the disorder in the material when constructively affecting the arrangement of the atomic lattice, so the actual disorder number, $n_{dis}$ will no longer equal $n_{imp}$ , requiring (91) be modified to

$$\gamma_{rp} = \frac{1}{1 + n_{dis}R^{V}_{s,\,imp;\,CP}} \qquad (95)$$

In this case, and relative to a nominal value, $n_{dis}$ may go negative, and the power modification exponential, will increase, if $R^{V}_{s,\,imp;\,CP}$ is positive. As a consequence, the critical temperature $T_c$ will decrease. However, if $R^{V}_{s,\,imp;\,CP}$ is negative, critical temperature $T_c$ will increase.

When modifying second term in the (95) denominator is considerably smaller than unity, (83), (84) and (95) allow $T_c$ to be written as

$$T_c = T_c^{BCS}\left(T_c\big|_{\substack{BCS\\ \exp}}\right)^{-n_{dis}R} \quad ; \quad T_c^{BCS} = \frac{\alpha}{\pi}\frac{\hbar\omega_D}{k_B}T_c\big|_{\substack{BCS\\ \exp}} \quad ; \quad T_c\big|_{\substack{BCS\\ \exp}} = e^{-\frac{2}{V'^{RPA,\,CP}_{eff,\,av}}\frac{1}{d(\varepsilon_F)}} \qquad (96)$$

The result in (96) is sketched in Figure 1, where $R$ stands for $R^{V}_{s,\,imp;\,CP}$ which is the impurity to electron-phonon renormalized potential energy ratio, and $T_c$ is plotted against $n_{dis}$. Notice, negative going values of $n_{dis}$ increase $T_c$ whereas positive going values decrease $T_c$. Parametrization in terms of $R$ is done. Note that in no cases may $n_{dis}$ < 0 be used to generate a superconducting effect when none existed in the material ($n_{dis}$ = 0 prior to disturbing the material), as that would be unphysical.

One may obtain a linearized version of (96) using a Taylor's expansion, such that

$$T_c(n_{dis}) = T_c^{(0)} + T_c^{(1)}n_{dis} + O([n_{dis}]^2) \qquad (97)$$

When this is done, $T_c^{(0)}$ and $T_c^{(1)}$ are determined to be

$$T_c^{(0)} = \frac{\alpha}{\pi}\frac{\hbar\omega_c}{k_B}e^{-\frac{1}{V'^{RPA,\,CP}_{eff,\,av}d(\varepsilon_F)/2}} \quad , \quad T_c^{(1)} = T_c^{(0)}\frac{2R^{V}_{s,\,imp;\,CP}}{V'^{RPA,\,CP}_{eff,\,av}d(\varepsilon_F)} \qquad (98)$$

Appearance of a finite $T_c^{(1)}$ coefficient is consistent with using a form of the Eliashberg equation employing the phonon spectral density $\alpha^2(\omega)F(\omega)$, finding additional low frequency transverse while reduced longitudinal electron-phonon coupling, giving a net effective gain in weighting by the phonon spectral density. This yields a critical temperature shift of $\left(T_c - T_c^p\right)/T_c^p = C_p/(q_D\ell_{mfp})$ [ Keck and Schmid, 1976]. Here $q_D$ is the Debye momentum (related to Debye temperature), $\ell_{mfp}$ the mean free path of electrons, and $T_c^p$ the critical temperature for pure metals, with $T_c$ the critical temperature for non-magnetic impurities. Arguing that $\ell_{mfp} \propto 1/n_{imp}$, yields a $T_c^{(1)} \propto n_{imp}$.

## XII. RELATING THE DISORDER POTENTIAL ENERGY TO THE GAP PARAMETER

The gap parameter used in the previous sections is not the regular or ordinary gap parameter. Rather, it has been modified by the impurity scattering potential energy, and so that altered gap is derived here. There are less typical or expected quantum many-body Green's functions, often referred to as anomalous Green's functions. One of them which will be of use is defined as

$$F_{\mathbf{k}\downarrow\uparrow}(\mathbf{k},\tau) \;=\; -\left\langle T_\tau\left\{c^\dagger_{-\mathbf{k}\downarrow}(\tau)c^\dagger_{\mathbf{k}\uparrow}(0)\right\}\right\rangle \qquad (99)$$

$T_\tau$ is the time ordering operator in the imaginary Matsubara time $\tau$ frame. The bracketed operation indicates an ensemble statistical average over any complete eigenstate set $v$. Thus for some operator $O$,

$$\langle O\rangle \;=\; \frac{1}{Z}\sum_v \langle v|O|v\rangle e^{-\beta E_v} \quad ; \quad Z \;=\; \sum_v e^{-\beta E_v} \qquad (100)$$

Here Z is the partition function, formed by the exponential sum using the system Hamiltonian eigenenergies $E_v$. The signature property of the anomalous Green's function is that it does not have mixed raising and lowering operators as the ordinary Green's function

$$G_{\uparrow\uparrow}(\mathbf{k},\tau) \;=\; -\left\langle T_\tau\left\{c_{\mathbf{k}\uparrow}(\tau)c^\dagger_{\mathbf{k}\uparrow}(0)\right\}\right\rangle \qquad (101)$$

These two types of Green's functions can be related by equations of motion,

$$\partial_\tau G_{\uparrow\uparrow}(\mathbf{k},\tau) \;=\; -\delta(\tau) \;-\; \xi_{\mathbf{k}} G_{\uparrow\uparrow}(\mathbf{k},\tau) \;+\; \Delta_{\mathbf{k}} F_{\downarrow\uparrow}(\mathbf{k},\tau) \qquad (102a)$$

$$\partial_\tau F_{\downarrow\uparrow}(\mathbf{k},\tau) \;=\; -\; \xi_{\mathbf{k}} F_{\downarrow\uparrow}(\mathbf{k},\tau) \;+\; \Delta_{\mathbf{k}} G_{\uparrow\uparrow}(\mathbf{k},\tau) \qquad (102b)$$

It is worth noting that the 2$^{nd}$ quantized raising and lowering electron operators employed in (99) and (101) are perturbed many-body operators.

A mean-field approach relates the total potential energy to the gap parameter using 2$^{nd}$ quantized electron operators. From it we need to find actual relations which explicitly demonstrate the relationship between the disorder effect via the impurity scattering potential energy, to the gap parameter. The starting point is recognizing that the expression uses the anomalous Green's function, converting into the Matsubara frequency domain, retrieving the 4-momentum space formula for it, eventually performing an integration.

$$
\begin{aligned}
\Delta_{\mathbf{k}} &= \sum_{\mathbf{k}'}^{|\xi_{\mathbf{k}'}|<\omega_D} V'_{\mathbf{k}\mathbf{k}'} \left\langle c_{-\mathbf{k}'\downarrow} c_{\mathbf{k}\uparrow} \right\rangle \\
&\approx V'^{RPA}_{eff,\,av} \sum_{\mathbf{k}'}^{|\xi_{\mathbf{k}'}|<\omega_D} F^{*}_{\downarrow\uparrow}(\mathbf{k},\tau)\Big|_{\tau\to 0} \\
&= V'^{RPA}_{eff,\,av} \sum_{\mathbf{k}}^{|\xi_{\mathbf{k}'}|<\omega_D} \sum_{ik_n} e^{-ik_n\tau} F^{*}_{\downarrow\uparrow}(\mathbf{k}, ik_n)\Big|_{\tau\to 0} \\
&= V'^{RPA}_{eff,\,av} \sum_{\mathbf{k}}^{|\xi_{\mathbf{k}'}|<\omega_D} \sum_{ik_n} e^{-ik_n\cdot 0^+} F^{*}_{\downarrow\uparrow}(\mathbf{k}, ik_n) \\
&= V'^{RPA}_{eff,\,av} \sum_{\mathbf{k}}^{|\xi_{\mathbf{k}'}|<\omega_D} \sum_{ik_n} e^{-ik_n\cdot 0^+} \left[ \frac{-\Delta^{*}_{\mathbf{k}}}{(ik_n)^2 - (E_{\mathbf{k}})^2} \right]^{*} \\
&= -\, V'^{RPA}_{eff,\,av} \sum_{\mathbf{k}}^{|\xi_{\mathbf{k}}|<\omega_D} \Delta_{\mathbf{k}} \frac{1}{\beta} \sum_{ik_n} e^{-ik_n\cdot 0^+} \frac{1}{(ik_n)^2 - (E_{\mathbf{k}})^2} \\
&= -\, V'^{RPA}_{eff,\,av} \sum_{\mathbf{k}}^{|\xi_{\mathbf{k}}|<\omega_D} \Delta_{\mathbf{k}} \frac{1}{\beta} \sum_{ik_n} e^{-ik_n\cdot 0^+} \frac{1}{(ik_n - |E_{\mathbf{k}}|)(ik_n + |E_{\mathbf{k}}|)}
\end{aligned}
\qquad (103)
$$

$\Delta_{\mathbf{k}}$ is readily evaluated by using the contour integration property for the complex valued function $g_0(z)$ with poles in the complex z-plane,

$$
\frac{1}{\beta} \sum_{ik_n} g_0(ik_n) e^{ik_n\tau} \Bigg|_{\tau>0} = \sum_{j=1}^{M} f(z_j) e^{z_j\tau} \operatorname*{Res}_{z=z_j}\left[ g_0(z) \right] \qquad (104)
$$

where $f$ is just the Fermi-Dirac function. Noting in (103) we used the following for $F_{\downarrow\uparrow}(\mathbf{k},\tau)$ and identify $g_0(z)$ as

$$
F_{\downarrow\uparrow} = \frac{-\Delta^{*}_{\mathbf{k}}}{(ik_n)^2 - (E_{\mathbf{k}})^2} \quad ; \quad g_0(z) = \frac{1}{(z - |E_{\mathbf{k}}|)(z + |E_{\mathbf{k}}|)} \qquad (105)
$$

one finds,

$$
\begin{aligned}
\Delta_{\mathbf{k}} &= -\, V'^{RPA}_{eff,\,av} \sum_{\mathbf{k}}^{|\xi_{\mathbf{k}'}|<\omega_D} \Delta_{\mathbf{k}} \frac{f(\beta E^{+}_{\mathbf{k}}) - f(\beta E^{-}_{\mathbf{k}})}{2E^{+}_{\mathbf{k}}} \\
&= -\, V'^{RPA}_{eff,\,av} \sum_{\mathbf{k}}^{|\xi_{\mathbf{k}'}|<\omega_D} \Delta_{\mathbf{k}} \frac{f(\beta|E_{\mathbf{k}}|) - f(-\beta|E_{\mathbf{k}}|)}{2|E_{\mathbf{k}}|} \\
&= -\, V'^{RPA}_{eff,\,av} \sum_{\mathbf{k}}^{|\xi_{\mathbf{k}'}|<\omega_D} \Delta_{\mathbf{k}} \frac{f(\beta E_{\mathbf{k}}) - f(-\beta E_{\mathbf{k}})}{2E_{\mathbf{k}}} \\
&= V'^{RPA}_{eff,\,av} \sum_{\mathbf{k}}^{|\xi_{\mathbf{k}'}|<\omega_D} \Delta_{\mathbf{k}} \frac{1 - 2f(\beta E_{\mathbf{k}})}{2E_{\mathbf{k}}}
\end{aligned}
\qquad (106)
$$

where we dropped the magnitude signs on the dispersion r elation, taking by default the positive branch.

Considering the simplest condition where the gap parameter becomes momentum independent, $\Delta_{\mathbf{k}} \to \Delta$, will factor out of the summation, leaving it to be divided out,

$$
\begin{aligned}
1 &= V'^{RPA}_{eff,\,av} \sum_{\mathbf{k}}^{|\xi_{\mathbf{k}'}|<\omega_D} \frac{1 - 2f(\beta E_{\mathbf{k}})}{2E_{\mathbf{k}}} \\
&= V'^{RPA}_{eff,\,av} \int_{-\varepsilon_F}^{\infty} D(\xi_{\mathbf{k}}) \frac{1 - 2f(\beta E_{\mathbf{k}})}{2E_{\mathbf{k}}} d\xi_{\mathbf{k}} \\
&\approx V'^{RPA}_{eff,\,av} D(\varepsilon_F) \int_{-\varepsilon_F}^{\infty} \frac{1 - 2f(\beta E_{\mathbf{k}})}{2E_{\mathbf{k}}} d\xi_{\mathbf{k}} \\
&\approx V'^{RPA}_{eff,\,av} D(\varepsilon_F) \int_{-\omega_D}^{\omega_D} \frac{1 - 2f(\beta E_{\mathbf{k}})}{2E_{\mathbf{k}}} d\xi_{\mathbf{k}} \\
&\approx V'^{RPA}_{eff,\,av} D(\varepsilon_F) \int_{-\omega_D}^{\omega_D} \frac{\tanh\left(\frac{\beta}{2}\sqrt{(\xi_{\mathbf{k}})^2 + |\Delta|^2}\right)}{2\sqrt{(\xi_{\mathbf{k}})^2 + |\Delta|^2}} d\xi_{\mathbf{k}}
\end{aligned} \tag{107}
$$

The final equality in (107) provides the relationship between the average effective RPA potential energy, as given by (85) and (86), and the gap parameter. [Note that $d(\varepsilon_F) = d_{st}(\varepsilon_F) = D(\varepsilon_F)/\mathcal{V}$, where $\mathcal{V}$ is the volume in 3D, which is replaced by area $A_{2D}$ in 2D.]

A non-transcendental relationship is found as $T \to 0$,

$$
|\Delta(0)| \approx |\Delta(0)|_{\substack{BCS\\pre}} \left( |\Delta(0)|_{\substack{BCS\\\exp}} \right)^{\gamma_{rp}} \tag{108}
$$

where

$$
|\Delta(0)|_{\substack{BCS\\pre}} = 2\hbar\omega_D \quad ; \quad |\Delta(0)|_{\substack{BCS\\\exp}} = e^{-\frac{1}{V'^{RPA,\,CP}_{eff,\,av} D(\varepsilon_F)}} \tag{109}
$$

As we saw earlier when examining the behavior of the critical temperature $T_c$, with a positive ratio , the power modification exponential will be less than unity, and applied to the $|\Delta(0)|_{\substack{BCS\\\exp}}$ factor, which is a unitless number less than one, the new effective BCS exponential will increased in size. This will have the effect of increasing $|\Delta(0)|$. On the other hand, if the ratio is negative, the power modification exponential will be greater than unity, and when applied to the $|\Delta(0)|_{\substack{BCS\\\exp}}$ factor, which is a unitless number less than one, since it scales down the temperature prefactor $T_c^{\alpha}\big|_{\substack{BCS\\pre}}$, the new effective BCS exponential will decreased in size. This will have the effect of decreasing $|\Delta(0)|$.

There is another subtlety to interpreting (87), and that is, increasing the number of impurities may reduce the disorder in the material when constructively affecting the arrangement of the atomic lattice, so the actual disorder number, $n_{dis}$ will no longer equal $n_{imp}$. In this case and relative to a nominal value, $n_{dis}$ may go negative, and (91) changes into (95), making $\gamma_{rp} = 1/\left[1 + n_{dis} R^{V}_{s,\,imp;\,CP}\right]$ increase, enhancing the value of $|\Delta(0)|$.

When modifying second term in the (95) denominator is considerably smaller than unity, (83), (84) and (95) allow $T_c$ to be written as

$$\Delta(0) = \Delta_{BCS}(0)\left(\Delta(0)\Big|_{\substack{BCS\\ \exp}}\right)^{-\,n_{dis}R} \quad ; \quad \Delta_{BCS}(0) = 2\hbar\omega_D\ \Delta(0)\Big|_{\substack{BCS\\ \exp}} \quad ; \quad \Delta(0)\Big|_{\substack{BCS\\ \exp}} = e^{-\frac{2}{V'^{RPA,\,CP}_{eff,\,av}}\frac{1}{d(\varepsilon_F)}} \qquad (110)$$

The result in (110) is sketched in Figure 2, where $R$ stands for $R^{V}_{s,\,imp;\,CP}$ which is the impurity to electron-phonon renormalized potential energy ratio, and $\Delta(0)$ is plotted against $n_{dis}$. Notice, negative going values of $n_{dis}$ increase $\Delta(0)$ whereas positive going values decrease $\Delta(0)$. Parametrization in terms of $R$ is done.

## XIII. CONCLUSIONS

Disorder as it microscopically affects superconducting properties, has been delineated in the treatment given here. Useful, compact analytical expressions for two major parameters of interest, the critical temperature $T_c$ and the gap $\Delta$, are found. The indirect relationship between $T_c$ and gap parameter $\Delta$ is shown when obtaining $\Delta$'s dependence on the same disorder potential energy quantity as $T_c$. The treatment shown here opens up the possibility that disorder could possibly increase $T_c$, something not seen in the 1980s, but observed in some circumstances as mentioned previously in the Introduction here. Hints of that possibility were indicated earlier in [Keck and Schmid, 1976].

The results here are not inconsistent with the original BCS theory [Bardeen, Cooper, Schrieffer, 1957], although formulas provided here are derived in a more streamlined fashion, not necessarily relying on bandstructure symmetries originally utilized in either reciprocal **k**-space or spin index σ.

## ACKNOWLEDGEMENTS

This contribution arose while working on the project Lower Dimensional Materials for Naval Applications, involving our Electronics Science & Technology, Material Science & Technology, and Chemistry Divisions, with the Plasma Physics Division also participating as a collaborating division, at the Naval Research Laboratory, Washington, D.C. Interactions with all of the project researchers in these divisions, have informed the contents of the present work, over the course of the project, from October 2013 – October 2017. I mention with particular gratitude the many interesting discussions with Dr. Michael S. Osofsky of MSTD, on electron-electron interactions, superconductivity, metal-insulator transition, and many other topics related to 2D solid state material systems. Finally, I thank several unnamed researchers, whose suggestions have been incorporated in improving the manuscript, reflected also in some of the references selected.

Wang, Y. L., X. L. Wu, C.-C. Chen, C. M. Lieber. (1990). Enhancement of the Critical Current Density in Single-Crystal $Bi_2Sr_2CaCu_2O_8$ Superconductors by chemically induced disorder, *Proc. Natl. Acad. Sci.* **87**, 7058, Sept.

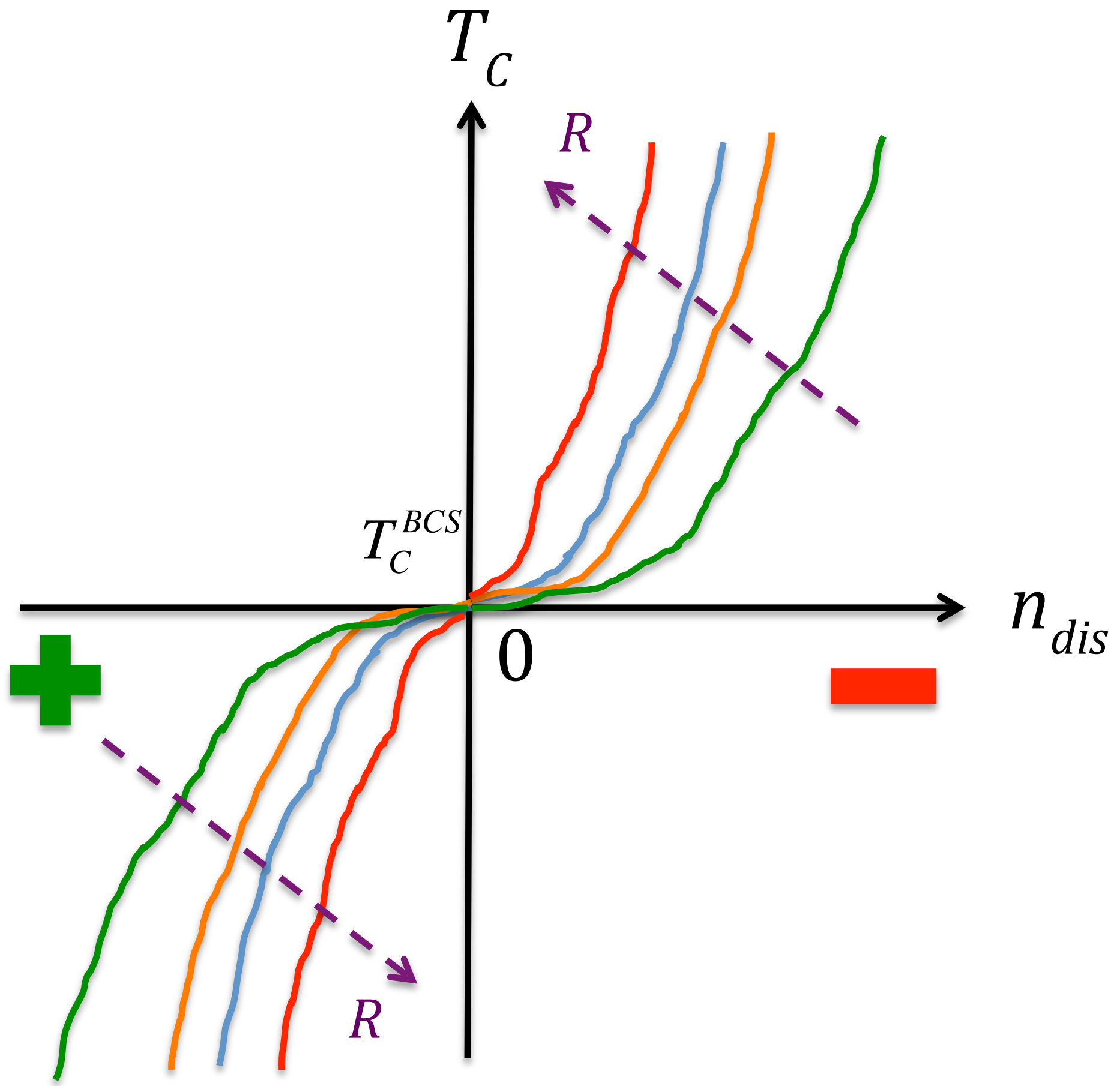

Figure 1: Critical superconducting temperature $T_c$ versus disorder density $n_{dis}$, parameterized in terms of electron-phonon renormalized energy ratio $R$.

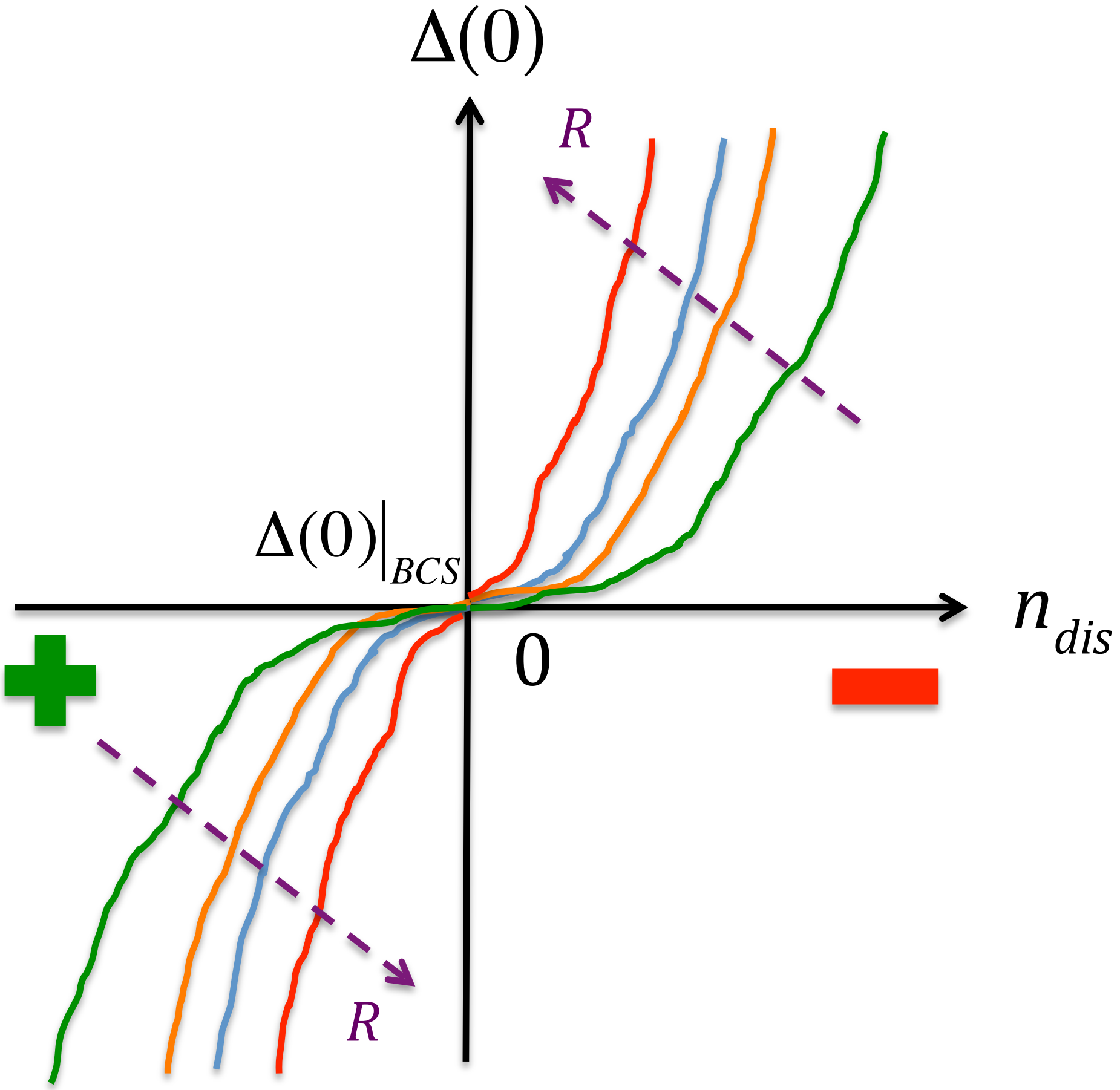

Figure 2: Gap Δ(0) at zero ambient temperature versus the disorder density $n_{dis}$, parameterized in terms of the electron-phonon renormalized energy ratio $R$.